\documentclass[journal]{IEEEtran}

\pdfoutput=1
\IEEEoverridecommandlockouts

\usepackage{acronym}
\usepackage{amsmath}
\usepackage{amssymb}
\usepackage[english]{babel}
\usepackage{caption}
\usepackage{cite}
\usepackage[shortlabels,inline]{enumitem}
\usepackage{etoolbox}
\usepackage{graphicx}
\usepackage{hyperref}
\usepackage{bookmark}
\usepackage{ifthen}
\usepackage{cleveref}
\usepackage[utf8]{inputenc}
\usepackage{lipsum}
\usepackage{siunitx}
\usepackage[dvipsnames]{xcolor}
\usepackage[T1]{fontenc}
\usepackage[subnum]{cases}
\usepackage{threeparttable}
\usepackage{tikz-cd}
\graphicspath{{./fig/}}
\usepackage[toc,symbols]{glossaries}
\usepackage{bm}

\allowdisplaybreaks

\newtheorem{remark}{Remark}
\newtheorem{theo}{Theorem}

\newtheorem{prop}{Proposition}
\newtheorem{assump}{Assumption}

\newcommand{\Figure}[1]{Fig.~\ref{#1}}
\newcommand{\Figures}[2]{Figs.~\ref{#1} and~\ref{#2}}

\newcommand{\Equation}[1]{\eqref{#1}}
\newcommand{\Equations}[2]{\eqref{#1} and~\eqref{#2}}
\newcommand{\EquationsList}[2]{\eqref{#1} --~\eqref{#2}}
\newcommand{\Table}[1]{Table~\ref{#1}}

\newcommand{\Section}[1]{Section~\ref{#1}}

\newcommand{\Sections}[2]{Sections~\ref{#1}~and~\ref{#2}}
\newcommand{\SectionsThree}[3]{Sections~\ref{#1}~,~\ref{#2},~and~\ref{#3}}

\newcommand{\Appendix}[1]{Appendix~\ref{#1}}

\newcommand{\Assumption}[1]{\textit{Assumption~\ref{#1}}}

\newcommand{\dx}{\ensuremath{\mathrm{d}x}}

\newcommand{\dzr}{\ensuremath{\mathrm{d}z^{\scriptscriptstyle{\mathrm{RX}}}}}
\newcommand{\dzm}{\ensuremath{\mathrm{d}z^{\scriptscriptstyle{\mathrm{TX}}}}}

\newcommand\given[1][]{\:#1\vert\:}
\newcommand\prob[1]{\textnormal{Pr}\{{#1}\}}
\newcommand\fpdf[2][]{\textnormal{f}_{\mathrm{#1}}\left(#2\right)}

\newcommand{\binomial}[1]{{\ensuremath{\textnormal{Binom}\Big({#1}\Big)}}}

\DeclareMathOperator\erf{erf}

\acrodef{1D}[1-D]{one-dimensional}
\acrodef{2D}[2-D]{two-dimensional}
\acrodef{3D}[3-D]{three-dimensional}
\acrodef{MC}{molecular communication}
\acrodef{OOK}{ON-OFF keying}
\acrodef{ISI}{inter-symbol interference}
\acrodef{IUI}{inter-user interference}
\acrodef{IR}{impulse response}
\acrodef{ML}{maximum likelihood}
\acrodef{wlog}[w.l.o.g.]{without loss of generality}
\acrodef{BER}{bit error rate}
\acrodef{SER}{symbol error rate}
\acrodef{BSC}{Binary Symmetric Channel}
\acrodef{CDF}{cumulative density function}
\acrodef{UCA}{uniform concentration assumption}
\acrodef{AWGN}{Additive White Gaussian Noise}
\acrodef{PBS}{particle-based simulation}
\acrodef{MLSE}{maximum likelihood sequence estimator}
\acrodef{VE}{viterbi equalizer}
\acrodef{MLE}{maximum likelihood estimator}
\acrodef{CIR}{channel impulse response}
\acrodef{CIRs}{channel impulse responses}
\acrodef{SNR}{signal-to-noise ratio}
\acrodef{SDMA}{Space Division Multiple Access}
\acrodef{MCDMA}{Molecular Code Division Multiple Access}
\acrodef{ADMA}{Amplitude-Division Multiple Access}
\acrodef{MTDMA}{Molecular Time Division Multiple Access}
\acrodef{MDMA}{Molecular Division Multiple Access}
\acrodef{MIMO}{multiple-input multiple-output}
\acrodef{TDMA}{Time Division Multiple Access}
\acrodef{CDMA}{Code Division Multiple Access}
\acrodef{CSI}{channel state information}
\acrodef{TX}{transmitter}
\acrodef{TXs}{transmitters}
\acrodef{RX}{receiver}
\acrodef{RXs}{receivers}
\acrodef{ARE}{area rate efficiency}
\acrodef{w.r.t.}{with respect to}
\acrodef{GFPD}{green fluorescent protein Dreiklang}
\acrodef{EX}{eraser}
\acrodef{MoSK}{molecule shift keying}

\newglossaryentry{zaSys}{%
  name = {\ensuremath{z_\mathrm{a}^{\scriptscriptstyle{\mathrm{Sys}}}}},
  description={ start system},
  type= symbols
}

\newglossaryentry{zbSys}{%
  name = {\ensuremath{z_\mathrm{b}^{\scriptscriptstyle{\mathrm{Sys}}}}},
  description={ end system},
  type= symbols
}

\newglossaryentry{zaEX}{%
  name = {\ensuremath{z_\mathrm{a}^{\scriptscriptstyle{\mathrm{EX}}}}},
  description={ start eraser},
  type= symbols
}

\newglossaryentry{zbEX}{%
  name = {\ensuremath{z_\mathrm{b}^{\scriptscriptstyle{\mathrm{EX}}}}},
  description={ end eraser},
  type= symbols
}

\newglossaryentry{zaTX}{%
  name = {\ensuremath{z_\mathrm{a}^{\scriptscriptstyle{\mathrm{TX}}}}},
  description={ start transmitter},
  type= symbols
}

\newglossaryentry{zbTX}{%
  name = {\ensuremath{z_\mathrm{b}^{\scriptscriptstyle{\mathrm{TX}}}}},
  description={ end transmitter},
  type= symbols
}

\newglossaryentry{zaRX}{%
  name = {\ensuremath{z_\mathrm{a}^{\scriptscriptstyle{\mathrm{RX}}}}},
  description={ start receiver},
  type= symbols
}

\newglossaryentry{zbRX}{%
  name = {\ensuremath{z_\mathrm{b}^{\scriptscriptstyle{\mathrm{RX}}}}},
  description={ end receiver},
  type= symbols
}

\newglossaryentry{lambdaEX}{%
  name = {\ensuremath{\lambda_{\mathrm{in}}^{\scriptscriptstyle{\mathrm{EX}}}}},
  description={ wavelength EX},
  type= symbols
}

\newglossaryentry{lambdaTX}{%
  name = {\ensuremath{\lambda_{\mathrm{in}}^{\scriptscriptstyle{\mathrm{TX}}}}},
  description={ wavelength TX},
  type= symbols
}

\newglossaryentry{lambdaRXin}{%
  name = {\ensuremath{\lambda_{\mathrm{in}}^{\scriptscriptstyle{\mathrm{RX}}}}},
  description={ wavelength RX in},
  type= symbols
}

\newglossaryentry{lambdaRXout}{%
  name = {\ensuremath{\lambda_{\mathrm{out}}^{\scriptscriptstyle{\mathrm{RX}}}}},
  description={ wavelength RX out},
  type= symbols
}

\newglossaryentry{PowerEX}{%
  name = {\ensuremath{P_{\mathrm{in}}^{\scriptscriptstyle{\mathrm{EX}}}}},
  description={ Power EX},
  type= symbols
}

\newglossaryentry{PowerTX}{%
  name = {\ensuremath{P_{\mathrm{in}}^{\scriptscriptstyle{\mathrm{TX}}}}},
  description={ Power TX},
  type= symbols
}

\newglossaryentry{PowerRXin}{%
  name = {\ensuremath{P_{\mathrm{in}}^{\scriptscriptstyle{\mathrm{RX}}}}},
  description={ Power RX in},
  type= symbols
}

\newglossaryentry{PowerRXout}{%
  name = {\ensuremath{P_{\mathrm{out}}^{\scriptscriptstyle{\mathrm{RX}}}}},
  description={ Power RX out},
  type= symbols
}

\newglossaryentry{VolumeEX}{%
  name = {\ensuremath{\mathrm{S}^{\scriptscriptstyle{\mathrm{EX}}}}},
  description={ Volume EX},
  type= symbols
}

\newglossaryentry{VolumeTX}{%
  name = {\ensuremath{\mathrm{S}^{\scriptscriptstyle{\mathrm{TX}}}}},
  description={ Volume TX},
  type= symbols
}

\newglossaryentry{VolumeTXNot}{%
  name = {\ensuremath{\smash{\overline{\mathrm{S}}}^{\scriptscriptstyle{\mathrm{TX}}}}},
  description={ Not Volume TX},
  type= symbols
}

\newglossaryentry{VolumeRX}{%
  name = {\ensuremath{\mathrm{S}^{\scriptscriptstyle{\mathrm{RX}}}}},
  description={ Volume RX},
  type= symbols
}

\newglossaryentry{AreaEX}{%
  name = {\ensuremath{A^{\scriptscriptstyle{\mathrm{EX}}}}},
  description={ Area EX},
  type= symbols
}

\newglossaryentry{AreaTX}{%
  name = {\ensuremath{A^{\scriptscriptstyle{\mathrm{TX}}}}},
  description={ Area TX},
  type= symbols
}

\newglossaryentry{AreaRX}{%
  name = {\ensuremath{A^{\scriptscriptstyle{\mathrm{RX}}}}},
  description={ Area RX},
  type= symbols
}

\newglossaryentry{VolumeSizeEX}{%
  name = {\ensuremath{V^{\scriptscriptstyle{\mathrm{EX}}}}},
  description={ Volume Size EX},
  type= symbols
}

\newglossaryentry{VolumeSizeTX}{%
  name = {\ensuremath{V^{\scriptscriptstyle{\mathrm{TX}}}}},
  description={ Volume Size TX},
  type= symbols
}

\newglossaryentry{VolumeSizeRX}{%
  name = {\ensuremath{V^{\scriptscriptstyle{\mathrm{RX}}}}},
  description={ Volume Size RX},
  type= symbols
}

\newglossaryentry{VolumeSizeSys}{%
  name = {\ensuremath{V^{\scriptscriptstyle{\mathrm{Sys}}}}},
  description={ Volume Size System},
  type= symbols
}

\newglossaryentry{lengthEX}{%
  name = {\ensuremath{l^{\scriptscriptstyle{\mathrm{EX}}}}},
  description={ Length EX},
  type= symbols
}

\newglossaryentry{lengthTX}{%
  name = {\ensuremath{l^{\scriptscriptstyle{\mathrm{TX}}}}},
  description={ Length TX},
  type= symbols
}

\newglossaryentry{lengthRX}{%
  name = {\ensuremath{l^{\scriptscriptstyle{\mathrm{RX}}}}},
  description={ Length RX},
  type= symbols
}

\newglossaryentry{lengthSys}{%
  name = {\ensuremath{L^{\scriptscriptstyle{\mathrm{Sys}}}}},
  description={ Length System},
  type= symbols
}

\newglossaryentry{NSys}{%
  name = {\ensuremath{N^{\scriptscriptstyle{\mathrm{Sys}}}}},
  description={ Signaling molecules in system},
  type= symbols
}

\newglossaryentry{CSys}{%
  name = {\ensuremath{C^{\scriptscriptstyle{\mathrm{Sys}}}}},
  description={ Concentration of signaling molecules in system},
  type= symbols
}

\newglossaryentry{TEX}{%
  name = {\ensuremath{T_{\mathrm{S}}^{\scriptscriptstyle{\mathrm{EX}}}}},
  description={ Irradiation time at EX},
  type= symbols
}

\newglossaryentry{TTX}{%
  name = {\ensuremath{T_{\mathrm{S}}^{\scriptscriptstyle{\mathrm{TX}}}}},
  description={ Irradiation time at TX},
  type= symbols
}

\newglossaryentry{TRX}{%
  name = {\ensuremath{T^{\scriptscriptstyle{\mathrm{RX}}}}},
  description={ Read out time at RX},
  type= symbols
}

\newglossaryentry{epsilonEX}{%
  name = {\ensuremath{\epsilon^{\scriptscriptstyle{\mathrm{EX}}}}},
  description={ Molar absorption coefficient EX},
  type= symbols
}

\newglossaryentry{epsilonTX}{%
  name = {\ensuremath{\epsilon^{\scriptscriptstyle{\mathrm{TX}}}}},
  description={ Molar absorption coefficient TX},
  type= symbols
}

\newglossaryentry{epsilonRX}{%
  name = {\ensuremath{\epsilon^{\scriptscriptstyle{\mathrm{RX}}}}},
  description={ Molar absorption coefficient RX},
  type= symbols
}

\newglossaryentry{PhotonInAbs}{%
  name = {\ensuremath{\Phi_{\mathrm{in,abs}}^{\scriptscriptstyle{m}}}},
  description={Photons into m which are absorbed},
  type= symbols
}

\newglossaryentry{PhotonInZero}{%
  name = {\ensuremath{\Phi_{\mathrm{in,}0}^{\scriptscriptstyle{m}}}},
  description={Photons into m which are absorbed},
  type= symbols
}

\newglossaryentry{PhotonInEX}{%
  name = {\ensuremath{\Phi_{\mathrm{in,abs}}^{\scriptscriptstyle{\mathrm{EX}}}}},
  description={ Photons into EX which are absorbed},
  type= symbols
}

\newglossaryentry{PhotonInTX}{%
  name = {\ensuremath{\Phi_{\mathrm{in,abs}}^{\scriptscriptstyle{\mathrm{TX}}}}},
  description={ Photons into TX which are absorbed},
  type= symbols
}

\newglossaryentry{PhotonInRX}{%
  name = {\ensuremath{\Phi_{\mathrm{in,abs}}^{\scriptscriptstyle{\mathrm{RX}}}}},
  description={ Photons into RX which are absorbed},
  type= symbols
}

\newglossaryentry{PhotonOutRX}{%
  name = {\ensuremath{\Phi_{\mathrm{out}}^{\scriptscriptstyle{\mathrm{RX}}}}},
  description={ Photons out of RX},
  type= symbols
}

\newglossaryentry{EnergyPhotonEX}{
  name = {\ensuremath{E_{\mathrm{in}}^{\scriptscriptstyle{\mathrm{EX}}}}},
  description={ Photons Energy of photons in EX},
  type= symbols
}

\newglossaryentry{EnergyPhotonRX}{
  name = {\ensuremath{E_{\mathrm{in}}^{\scriptscriptstyle{\mathrm{RX}}}}},
  description={ Photons Energy of photons in RX},
  type= symbols
}

\newglossaryentry{prPhi}{
  name = {\ensuremath{p_{\Phi}}},
  description={ quantum yield},
  type= symbols
}

\newglossaryentry{prAStern}{
  name = {\ensuremath{p_{\mathrm{A}^*}}},
  description={excited A molecules fraction},
  type= symbols
}

\newglossaryentry{prDet}{
  name = {\ensuremath{p_{\mathrm{d}}}},
  description={detected at RX which is prAstern x prPhi},
  type= symbols
}

\newglossaryentry{prSwitchTX}{
  name = {\ensuremath{p_{\mathrm{s}}^{\scriptscriptstyle{\mathrm{TX}}}}},
  description={ switching probability in TX},
  type= symbols
}

\newglossaryentry{prSwitchSpon}{
  name = {\ensuremath{p_{\mathrm{s,\scriptscriptstyle{SP}}}^{\scriptscriptstyle{\mathrm{RX}}}}},
  description={ Spontaneous swithcing probability},
  type= symbols
}

\newglossaryentry{prTX}{
  name = {\ensuremath{p^{\scriptscriptstyle{\mathrm{TX}}}}},
  description={ Probability to be in TX at sampling time},
  type= symbols
}

\newglossaryentry{prSwitchEX}{
  name = {\ensuremath{p_{\mathrm{s}}^{\scriptscriptstyle{\mathrm{EX}}}}},
  description={Switching probability in EX},
  type= symbols
}

\newglossaryentry{prAEX}{
  name = {\ensuremath{p_{\mathrm{A}}^{\scriptscriptstyle{\mathrm{EX}}}}},
  description={ Probability of state A at entrance of EX},
  type= symbols
}

\makeatletter
\long\def\@makecaption#1#2{\ifx\@captype\@IEEEtablestring%
    \footnotesize\begin{center}{\normalfont\footnotesize #1}\\
        {\normalfont\footnotesize\scshape #2}\end{center}%
    \@IEEEtablecaptionsepspace
    \else
    \@IEEEfigurecaptionsepspace
    \setbox\@tempboxa\hbox{\normalfont\footnotesize {#1.}~~ #2}%
    \ifdim \wd\@tempboxa >\hsize%
    \setbox\@tempboxa\hbox{\normalfont\footnotesize {#1.}~~ }%
    \parbox[t]{\hsize}{\normalfont\footnotesize \noindent\unhbox\@tempboxa#2}%
    \else
    \hbox to\hsize{\normalfont\footnotesize\hfil\box\@tempboxa\hfil}\fi\fi}
\makeatother

\renewcommand{\arraystretch}{1}
\newcommand{\scaleSection}{\vspace{0cm}}
\newcommand{\scaleSubsection}{\vspace{0cm}}
\newcommand{\scaleSubsubsection}{\vspace{0cm}}
\newcommand{\scaleSectionBelow}{\vspace{0cm}}
\newcommand{\scaleSubsectionBelow}{\vspace{0cm}}
\newcommand{\scaleSubsubsectionBelow}{\vspace{0cm}}
\newcommand{\scaleAlign}{\vspace{0cm}}

\begin{document}
\bstctlcite{disable_url}
\title{Media Modulation based Molecular \\ Communication
}
\author{
\IEEEauthorblockN{Lukas Brand, Moritz Garkisch, Sebastian Lotter, Maximilian Schäfer, Andreas \\Burkovski, Heinrich Sticht, Kathrin Castiglione, and Robert Schober}\thanks{This manuscript was presented in part at the 2022 IEEE International Conference on Communications \cite{brand2021media}.}
}

\maketitle
\begin{abstract}
In conventional molecular communication (MC) systems, the signaling molecules used for information transmission are stored, released, and then replenished by a transmitter (TX). However, the replenishment of signaling molecules at the TX is challenging in practice. Furthermore, in most envisioned MC applications, e.g., in the medical field, it is not desirable to insert the TX into the MC system, as this might impair natural biological processes. In this paper, we propose the concept of media modulation based MC where the TX is placed outside the channel and utilizes signaling molecules already present inside the system. The signaling molecules can assume different states which can be switched by external stimuli. Hence, in media modulation based MC, the TX modulates information into the state of the signaling molecules. In particular, we exploit the group of photochromic molecules, which undergo light-induced reversible state transitions, for media modulation. We study the usage of these molecules for information transmission in a three-dimensional duct system, which contains an eraser, a TX, and a receiver for erasing, writing, and reading of information via external light, respectively. We develop a statistical model for the received signal which accounts for the distribution of the signaling molecules in the system, the initial states of the signaling molecules, the reliability of the state control mechanism, the randomness of irrepressible, spontaneous state switching, and the randomness of molecule propagation. We adopt a maximum likelihood detector and show that it can be reduced to a threshold based detector. Furthermore, we derive analytical expressions for the optimal threshold value and the resulting bit error rate (BER), respectively. Both the statistical model and BER results are verified by computer simulations. Our results reveal that media modulation enables reliable information transmission, validating it as a promising alternative to MC based on molecule emitting TXs.  

\end{abstract}

\setlength{\belowdisplayskip}{2pt}
\setlength{\belowdisplayshortskip}{2pt}
\acresetall
\scaleSection
\section{Introduction}\label{intro}
\scaleSectionBelow
To facilitate synthetic \ac{MC}, several modulation schemes for embedding information into molecular signals have been proposed over the last few years \cite{kuran2020survey}. In \cite{Kuran2011}, concentration shift keying (CSK) and \ac{MoSK}, where information is encoded into the molecule concentration and the molecule type, respectively, have been proposed. Additionally, index modulation techniques such as molecular space shift keying \cite{Huang2019SpatMod, Gursoy2019IndMod} have been reported for \ac{MC} systems with multiple spatially distributed \ac{TX} locations, where information is embedded into the activation of a single transmitter.
Moreover, the authors in \cite{Tang2021} proposed molecular type permutation shift keying (MTPSK) where information is encoded into permutations of different molecule types.
\par
For the implementation of these modulation schemes, the signaling molecules are stored, released, and then replenished by the transmitter.
However, from a practical point of view, the replenishment of signaling molecules at the TX is difficult to realize, especially at microscale and in medical applications.
Therefore, it is desirable to develop alternative concepts for both \ac{TX} and modulation design that decouple the modulation process from the release of signaling molecules. \par
Such a decoupling has already been proposed for conventional wireless communication systems in the context of media modulation.
Here, information is embedded into the properties of the communication medium, i.e., the carrier signal is not directly modulated \cite{Khandani2013, Basar2019}.
Extending the concept of media modulation to MC, the authors of \cite{gohari2016information} propose to alter the properties of the channel to embed information. In \cite{farahnak2020molecular}, changing the flow velocity of the medium for modulation is proposed. In contrast to the channel-based forms of media modulation considered in \cite{gohari2016information, farahnak2020molecular}, in this paper, we propose a new form of media modulation, where the properties of signaling molecules already present in the channel are modulated for information transmission via an external stimulus.

In the simplest case, the signaling molecules can be in two different states which can be switched by an external stimulus to encode the information to be conveyed. For media modulation, the signaling molecules may be naturally present in the environment or they may be injected once and then remain in the channel.  Media modulation is different from existing modulation schemes as information is not encoded in the molecule concentration, the type of molecule or the releasing transmitter but in the properties of the signaling molecules themselves. Media modulation based MC can overcome several shortcomings of existing MC systems by avoiding repeated injections of signaling molecules, hereby (i) making the replenishment of the TX unnecessary\footnote{We note that if the signaling molecules degrade in the channel, occasional replenishment of signaling molecules may be necessary. However, even in this case, the number of additional signaling molecules needed is expected to be much lower than for a conventional \ac{MC} system with molecule emitting \ac{TX}, provided suitable signaling molecules are chosen.}, and (ii) reducing the soiling of the channel due to the deposition of signaling molecules after repeated injection.
Furthermore, the TX, i.e., the source of the external stimulus, is not placed inside the transmission medium and does not influence molecule propagation.
The practical feasibility of media modulation hinges on the availability of suitable switchable signaling molecules, of course.
The general concept of this form of media modulation based \ac{MC} is new and has not been reported in the literature, yet.
Nevertheless, redox-based MC \cite{Kim2019} can be interpreted as an instance of media modulation where signaling particles are switched between two states, the reduced state and the oxidized state.
However, the focus of \cite{Kim2019} was on the biological and experimental aspects of redox-based MC, while a thorough communication theoretical analysis was not conducted.
Furthermore, in biological systems, a natural form of media modulation can be observed during phosphorylation, where a phosphoryl group is added to a protein affecting the properties of the protein. The phosphorylation is mediated by a kinase, a specific type of enzyme, which in turn can be \mbox{controlled via an external stimulus \cite{grusch2014spatio, chang2014light}.}

Another promising candidate for signaling molecules for media modulation are photochromic molecules. These molecules can be reversibly interconverted between two states, where the transition between the states is induced by light \cite{balzani2014photochemistry}. Photochromic molecules are well established in molecular devices for information processing\cite[Chap.~10]{balzani2014photochemistry}, but their exploitation for media modulation in synthetic MC systems is new.
Photochromic systems facilitate different functionalities including writing, reading, and erasing of information by an external light stimulus.

In this paper, we develop a novel modulation scheme for synthetic MC systems employing photochromic signaling molecules.
Hereby, the information is embedded into the state of a photochromic molecule.
As photochromic signaling molecule, we exemplarily consider the reversibly photo-switchable green fluorescent protein variant ``Dreiklang'' (GFPD)\acused{GFPD}, whose fluorescence can be reversibly switched by light stimuli of mutually different wavelengths \cite{brakemann2011reversibly}. Fluorescence denotes the ability of a molecule to first absorb light and then radiate light back at a higher wavelength, i.e., with a lower energy, which makes it possible to read out the current state of a \ac{GFPD} molecule. \ac{GFPD} has been shown durable, i.e., stable over several photo-switching cycles \cite{brakemann2011reversibly}, which allows the long-term use of \ac{GFPD} as signaling molecule.

The application of fluorescence has gained significant interest for the design of synthetic MC systems over the last few years \cite{Damrath2021,Tuccitto2017,Amerizadeh2021,Krishnaswamy2013,Dadi2021,Kuscu2015}.
In the experimental MC system in \cite{Damrath2021}, a fluorescent dye was employed as information carrier.
In order to facilitate long-range information transfer in MC systems, fluorescent carbon quantum dots were investigated as signaling particles due to their concentration dependent emission properties \cite{Tuccitto2017}.
In \cite{Amerizadeh2021}, a bacterial receiver (RX)\acused{RX} was proposed which produces green fluorescent protein upon the reception of signaling molecules.
Similarly, the authors of \cite{Krishnaswamy2013} employed genetically engineered \textit{Escherichia coli} (\textit{E. coli}) bacteria to exhibit fluorescence upon the reception of specific signaling molecules, thereby demonstrating that a chemical transmit signal can be converted to a fluorescence signal at the RX.
In \cite{Dadi2021}, an MC system for generating pulse shaped signals with a NOR logic operation in engineered \textit{E.~coli} bacteria was proposed. Hereby, the fluorescence of the yellow fluorescent protein was employed as the pulse signal generated by the NOR gate.
The authors in \cite{Kuscu2015} developed a \ac{MIMO} nano-communication system, where information transmission was based on the exchange of energy levels between fluorescent molecules employed as transceiver nano-antennas.
Most of the existing works employ fluorescent molecules because of their easy detectability at the RX.
However, to the best of the authors' knowledge, the encoding of information into the state of a switchable photochromic molecule was not considered so far in the MC literature.

In the \ac{MC} system proposed in this paper, fluorescent and non-fluorescent GFPD molecules correspond to state A and state B, respectively.
The fluorescence of GFPD can be switched on (B$\to$A) and off (A$\to$B) by light stimuli at different wavelengths.
Consequently, assuming the GFPD molecules are suspended in a fluid medium as elements of the envisioned synthetic MC system, optical sources emitting light at different wavelengths can be used as TX unit to modulate information (B$\to$A) and \ac{EX} unit to delete the information (A$\to$B). Moreover, a fluorescence detector equipped with a fluorescence-stimulating light source and an optical sensor can be employed as RX.

The main contributions of this paper can be summarized as follows:
\begin{itemize}
	\item We propose a new form of media modulation for synthetic MC. As information carrier, we employ signaling molecules, which are already present in the channel. These molecules can assume different states which can be reversibly switched and read out by external stimuli. This property allows for information transmission by writing, reading, and erasing information embedded in the molecule state. Hence, this novel concept does not require a \ac{TX} that stores and releases molecules, and therefore overcomes several drawbacks of existing modulation schemes, e.g., the need for TX replenishment.
	\item We realize media modulation by employing photochromic molecules for information transmission. The information is conveyed in the state of these molecules, which is either fluorescent or non-fluorescent. Since both the state transitions and the fluorescence are triggered by external light stimuli, we provide an in-depth mathematical investigation of the underlying photochemical processes.
	\item We develop a statistical model for the received signal which integrates the different sources of randomness present in media modulation based MC systems.
	\item Based on the statistical model, we derive mathematical expressions for the optimal threshold value of a threshold-based detector and the \ac{BER}.
	\item We show that reliable media modulation based \ac{MC} is feasible and provide insight into the impact of the irradiation power densities, the number of signaling molecules present in the channel, and the channel characteristics, such as the flow velocity, on the \ac{BER}.
\end{itemize}
The proposed concept of media modulation was introduced in the conference version \cite{brand2021media} of this paper. Here, we significantly extend \cite{brand2021media} by investigating the physical properties of the \ac{EX} and the \ac{RX} in detail. Furthermore, we include the spontaneous state switching of \ac{GFPD} molecules in our model. The aforementioned extensions require the development of a more elaborate and comprehensive statistical model as compared to the model proposed in \cite{brand2021media}. Based on this new statistical model, we provide mathematical expressions for the optimal detection threshold of a threshold detector and the BER.

The remainder of this paper is organized as follows.
In Section~\ref{sec:model}, we introduce the considered MC system and describe the proposed media modulation scheme.
In Section~\ref{photoChemic}, the relevant photochemical processes are studied in detail.
In Section~\ref{sec:math}, we derive an analytical end-to-end statistical model for the proposed \ac{MC} system. Expressions for the detection threshold and the \ac{BER} are provided in Section~\ref{sec:performance}.
In Section~\ref{sec:evaluation}, we evaluate the proposed models numerically.
Section~\ref{sec:conclusion} concludes the paper and outlines topics for future work. Note that we use the terms \textit{switching} and \textit{conversion} interchangeably.

\scaleSection
\section{System Model}
\label{sec:model}
\begin{figure*}[!tbp]
  \centering
  \includegraphics[width=0.85\textwidth]{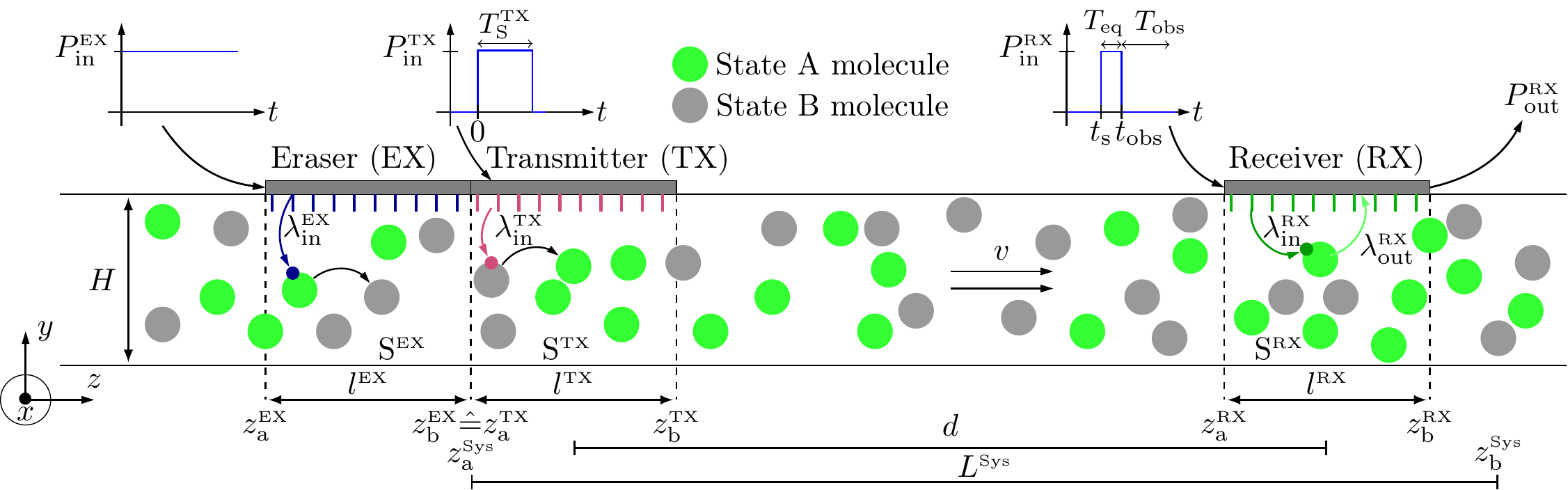}
  \caption{System model: Signaling molecules are uniformly distributed in a 3-D volume with extent $H$ in $y$-direction, $W$ in $x$-direction, and infinite extent in $z$-direction, resembling an infinite pipe. The molecules propagate by Brownian motion and uniform flow. When signaling molecules enter the \ac{EX}, their state is random (either A (green) or B (gray)). To reduce randomness, the EX switches the molecules in state A to state B via irradiation of wavelength $\gls{lambdaEX}$. For transmission of a binary $1$, the molecules are switched from state B to state A at the \ac{TX} via irradiation with wavelength $\gls{lambdaTX}$. The \ac{RX} triggers a fluorescence reaction in the state A molecules, which results in the emission of an optical signal with wavelength $\gls{lambdaRXout}$ and measurable light power density $\gls{PowerRXout}$. We refer to \Figures{block_diagram_TX_EX}{block_diagram_RX} for a detailed description of the photochemical reactions. The variables used in \Figure{system_model_picture} are defined in \Section{sec:model} and \Table{Table:Parameter}. \Table{Table:Parameter} also lists corresponding default values.}
  \label{system_model_picture}
\end{figure*}

\scaleSectionBelow
In this section, we describe the proposed media modulation based \ac{MC} system, including the erasure, modulation, propagation, and reception mechanisms, cf. \Figure{system_model_picture}. The system model presented in this section is generic and applicable to different types of photoswitchable fluorescent molecules. The mathematical details of the relevant photochemical processes are given in \Section{photoChemic}. In \Section{sec:evaluation}, the model is specialized to \ac{GFPD}.
\scaleSubsection
\subsection{Topology}\label{general_system_model}
\scaleSubsectionBelow
\Figure{system_model_picture} shows the considered \ac{3D} straight rectangular duct with height $H$, width $W$, and infinite extent in $z$-direction, which we denote as domain $\Omega$. The duct is filled with a fluid medium, which flows in $z$-direction with constant velocity $v>0$, i.e., we assume uniform flow, as is widely done in the \ac{MC} literature \cite{Jamali2019ChannelMF}, cf. \Section{propagation}. Moreover, we assume that the duct surface is impermeable to molecules, i.e., the molecules are reflected at the boundaries of the duct. Additionally, at duct segments $\gls{VolumeEX}$, $\gls{VolumeTX}$, and $\gls{VolumeRX}$, where the \ac{EX}, \ac{TX}, and \ac{RX} are located, respectively, we assume the duct surface to be transparent to light. The communication process of interest takes place in the subvolume $\mathrm{S} = \left\{\Omega \land \gls{zaSys} \leq z \leq \gls{zbSys}\right\}$ of length $\gls{lengthSys} = \gls{zbSys} - \gls{zaSys}$ with unit $\si{\meter}$ and volume size $\gls{VolumeSizeSys} = W H \gls{lengthSys}$ with unit $\si{\meter^3}$. In the following, all lengths and volume sizes have the units $\si{\meter}$ and $\si{\meter^3}$, respectively. We employ photochromic signaling molecules, which can assume two distinguishable states, state A and state B, and, depending on their state, we refer to them as state A molecules and state B molecules, respectively. The state of a photochromic molecule can be changed by irradiation with light at appropriate wavelength.


We assume that the state of the signaling molecules is random when they arrive at the \ac{EX} at $z = \gls{zaEX}$, i.e., with a probability of $\gls{prAEX} \in [0, 1]$ and $1-\gls{prAEX}$ a signaling molecule enters the \ac{EX} in state A and state B, respectively. We assume that $\gls{prAEX}$ is determined by an external process that cannot be controlled in the considered system.
At time $t=0$, the signaling molecules are assumed to be uniformly distributed in the duct system, i.e., we assume that the signaling molecules had enough time to reach an equilibrium \ac{w.r.t.} space. The latter assumption requires both identical propagation behavior of state A and state B molecules, cf. \Section{propagation}, and independence between the state switching processes and molecule propagation, cf. \Section{photoChemic}. We assume that $\gls{NSys}$ signaling molecules are uniformly distributed in subvolume S at time $t=0$, i.e., we assume that the concentration of the signaling molecules is $\gls{CSys} = \frac{\gls{NSys}}{\gls{VolumeSizeSys} N_{\mathrm{Av}}}$ with unit $\si{\mathrm{mol} \, \metre^{-3}}$, where $N_{\mathrm{Av}}$ denotes the Avogadro constant\footnote{The assumption that the number of signaling molecules in subvolume S is known is similar to the common assumption that the number of molecules released by the TX in conventional MC systems is exactly known \cite{Jamali2019ChannelMF}. In practice, only the average number of signaling molecules in subvolume S, $\mathbb{E}\{\gls{NSys}\}$, may be known. Extending the analysis in this paper to this case is an interesting topic for future work.}. In the following, all concentrations have the unit $\si{\mathrm{mol} \, \metre^{-3}}$.

\scaleSubsection
\subsection{Purification at the EX}\label{EX_details}
\scaleSubsectionBelow
We consider a \ac{2D} \ac{EX} with area $\gls{AreaEX} = \gls{lengthEX} W$, which is attached to the surface of the duct at $\gls{zaEX} \leq z \leq \gls{zbEX}$, i.e., the \ac{EX} length is $\gls{lengthEX} = \gls{zbEX} - \gls{zaEX}$, cf. \Figure{system_model_picture}. The \ac{EX} shall ensure that molecules entering subvolume S are in state B, as the \ac{TX} assumes all molecules are in state B. To this end, the \ac{EX} constantly radiates light of wavelength $\gls{lambdaEX}$ and power density $\gls{PowerEX}$ with unit $\si{\watt \per\m \squared}$ into volume $\gls{VolumeEX} = \left\{\Omega \land \gls{zaEX} \leq z \leq \gls{zbEX}\right\}$ of size $\gls{VolumeSizeEX} = H \gls{AreaEX}$. In the following, all power densities have the unit $\si{\watt \per\m \squared}$. The \ac{EX} radiation triggers a photochemical reaction at the signaling molecules that switches the state of a signaling molecule from state A to state B with probability $p_{\mathrm{s, EX}} \in [0,1]$.

\scaleSubsection
\subsection{Media Modulation at the TX}\label{media_modulation_TX}
\scaleSubsectionBelow
We consider a \ac{2D} \ac{TX} with area $\gls{AreaTX} = \gls{lengthTX} W$, which is attached to the surface of the duct next to the \ac{EX} at $\gls{zbEX} = \gls{zaTX} \leq z \leq \gls{zbTX}$, i.e., the \ac{TX} length is $\gls{lengthTX} = \gls{zbTX} - \gls{zaTX}$, cf. \Figure{system_model_picture}. For information transmission, at $t=0$, the \ac{TX} radiates light of wavelength $\gls{lambdaTX}$ and power density $\gls{PowerTX}$ for an irradiation duration of $\gls{TTX}$ into volume $\gls{VolumeTX} = \left\{\Omega \land \gls{zaTX} \leq z \leq \gls{zbTX}\right\}$ of size $\gls{VolumeSizeTX} = H \gls{AreaTX}$. The \ac{TX} uses \ac{OOK} modulation \cite{Jamali2019ChannelMF} and either radiates with power density $\gls{PowerTX} > 0$ or is inactive, i.e., $\gls{PowerTX} = 0$, representing binary symbols $s = 1$ and $s = 0$, respectively. Assuming appropriate source coding, the empirical frequency of data symbols $0$ and $1$ will approach $0.5$ in the limit of long data transmission, i.e., we assume that the binary information bits $0$ and $1$ are equiprobable. The \ac{TX} radiation triggers a photochemical reaction that switches the state of a signaling molecule from state B to state A with probability $\gls{prSwitchTX} \in [0,1]$.

We consider single symbol transmission, i.e., \ac{ISI} is assumed to be negligible. We make this assumption to focus on unveiling the properties of the novel modulation scheme proposed. Single symbol transmission can be realized by introducing a guard interval between symbols. The length of the guard interval has to be chosen in accordance with the system properties such that consecutive transmissions do not overlap, which may significantly reduce the achievable transmission rate. If a guard interval cannot be afforded, \ac{ISI} mitigation techniques as proposed in \cite{dambri2019performance}, where light-based eraser units at the \ac{RX} side are utilized, can be applied.

\scaleSubsection
\subsection{Detection at the RX}\label{general_RX}
\scaleSubsectionBelow
We consider a \ac{2D} \ac{RX} with area $\gls{AreaRX} = \gls{lengthRX} W$, which is attached to the surface of the duct at $\gls{zaRX} \leq z \leq \gls{zbRX}$, i.e., the \ac{RX} length is $\gls{lengthRX} = \gls{zbRX} - \gls{zaRX}$.  We assume that the \ac{TX} and the \ac{RX} are parallel to each other and therefore the distance between the \ac{TX} and \ac{RX} centers is $d = \frac{\gls{zbRX}+\gls{zaRX}}{2} - \frac{\gls{zbTX}+\gls{zaTX}}{2}$. The \ac{RX} emits light of wavelength $\gls{lambdaRXin}$ and can sense light of wavelength $\gls{lambdaRXout}$ emitted by the signaling molecules in state A. At fixed sampling time $t_{\mathrm{s}} = \frac{d}{v}$, the \ac{RX} radiates light of wavelength $\gls{lambdaRXin}$ and power density $\gls{PowerRXin}$ for a short time duration of $T_{\mathrm{eq}}$ into volume $\gls{VolumeRX} =\left\{\Omega \land \gls{zaRX} \leq z \leq \gls{zbRX}\right\}$ of size $\gls{VolumeSizeRX} = H \gls{AreaRX}$. Here, $t_\mathrm{s}$ corresponds to the time a molecule needs to propagate from \ac{TX} to \ac{RX} due to uniform flow. The radiated light triggers a fluorescence reaction in state A molecules, i.e., illuminated state A molecules radiate light with wavelength $\gls{lambdaRXout}$. This results in a light power density $\gls{PowerRXout}$ received at \gls{AreaRX}. We denote the probability to detect a state A molecule at the \ac{RX} by the abovementioned process as $\gls{prDet}$.
\begin{remark}
  All considered wavelengths $\gls{lambdaEX}$, $\gls{lambdaTX}$, $\gls{lambdaRXin}$, and $\gls{lambdaRXout}$ are mutually distinct. For \ac{GFPD}, the values of these wavelengths are provided in \Table{Table:Parameter}.
\end{remark}

\scaleSubsection
\subsection{Molecule Propagation}\label{propagation}
\scaleSubsectionBelow
In both states A and B, the molecules are subject to \ac{3D} Brownian motion characterized by diffusion coefficients $D_{\mathrm{A}}$ and $D_{\mathrm{B}}$, respectively, and flow. The actual flow for the given system structure is laminar, which is for simplicity commonly approximated by uniform flow \cite{Jamali2019ChannelMF}. Hence, we also assume uniform flow with constant flow velocity $v$ in this work. While in general a chemical reaction can result in a change of the molecule structure and size, here we assume $D = D_{\mathrm{A}} = D_{\mathrm{B}}$ for \ac{GFPD}.

\scaleSection
\section{Photochemical Processes}
\label{photoChemic}
\scaleSectionBelow
\begin{figure*}
    \centering
    \begin{minipage}[t]{0.48\textwidth}
      \centering
      \includegraphics[width=0.7\columnwidth]{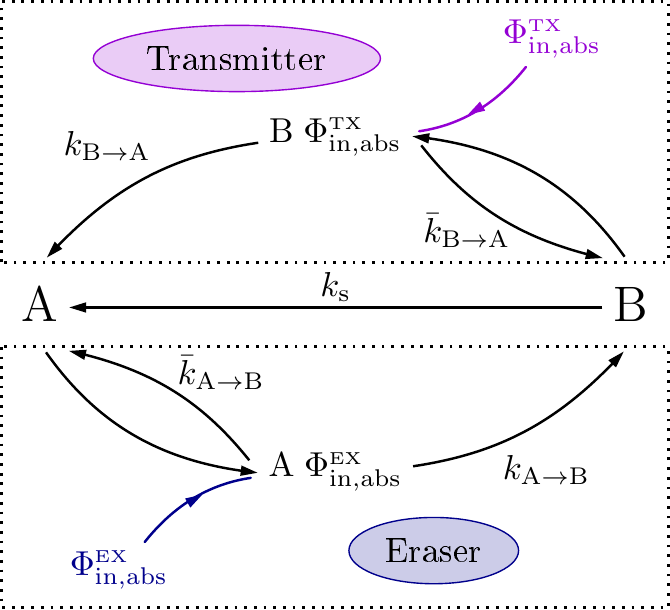}
      \caption{The photochemical reactions at \ac{EX} lower dotted box) and \ac{TX} upper dotted box) are shown with the relevant reaction rate constants. The different colors indicate the different wavelengths $\gls{lambdaEX}$ and $\gls{lambdaTX}$ of the involved photons $\gls{PhotonInEX}$ and $\gls{PhotonInTX}$, respectively. In addition, the spontaneous switching from state B molecule to state A molecule is shown by the black arrow with label $k_{\mathrm{{s}}}$. The loops between the states of the signaling molecules indicate that photochromic molecules are reversibly switchable.}\label{block_diagram_TX_EX}
    \end{minipage}
    \hfill
    \begin{minipage}[t]{0.48\textwidth}
      \centering
      \includegraphics[width=0.7\columnwidth]{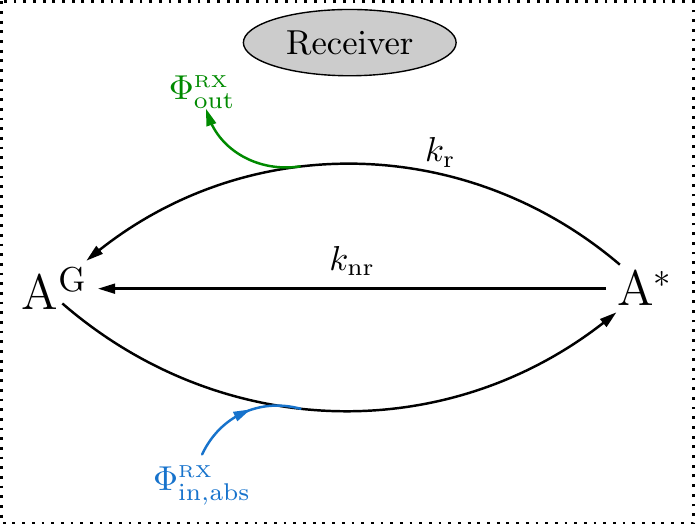}
      \caption{In the \ac{RX}, molecules in state A can assume the two substates $\mathrm{A}^*$ and $\mathrm{A}^{\mathrm{G}}$. The photochemical fluorescence process at the \ac{RX} is shown with the relevant reaction rate constants. The different colors indicate the wavelengths $\gls{lambdaRXin}$ and $\gls{lambdaRXout}$ of the involved photons $\gls{PhotonInRX}$ and $\gls{PhotonOutRX}$, respectively. The switching from $\mathrm{A}^*$ to $\mathrm{A}^{\mathrm{G}}$ without the radiation of a photon is shown by a black arrow with label $k_{\mathrm{{nr}}}$.}\label{block_diagram_RX}
    \end{minipage}
\end{figure*}

In the proposed system, we employ photochromic molecules as signaling molecules, which can be reversibly switched from state A to state B by external light. This property allows reading, writing, and erasing of information embedded in the state of a molecule by irradiation with light of different wavelengths.

In this section, we mathematically model the photochemical processes, which enable the state transitions and state detection of the signaling molecules. In \Section{general:photochemic}, we introduce the general properties and underlying mathematically descriptions of photochemical processes, upon which we build in \Section{section_photochemic_at_the_sections} to derive mathematical expressions for the probabilities of, respectively, state switching at the \ac{EX} and \ac{TX}, spontaneous state switching in the propagation channel, and detection at the \ac{RX} via fluorescence. We summarize in \Appendix{sections_assumptions} all simplifying assumptions made to arrive at tractable models and explain \mbox{the conditions needed for their validity.}

\scaleSubsection
\subsection{Quantitative Analysis of a Photoreaction: Absorption and Quantum Yield}
\label{general:photochemic}
\scaleSubsectionBelow

Any radiation of light corresponds to a photon flux $q_{w}^{m}$ with unit $\si{\per \second}$, which is given by
\scaleAlign
\begin{align}
  q_{w}^{m} = \frac{P_{w}^{m} A^{m} }{  E_{w}^{m}},
  \label{eq:Light_source_to_photons}
\end{align}
where $m \in \{ \mathrm{EX}, \mathrm{TX}, \mathrm{RX}\}$ and $w \in \{\mathrm{in}, \mathrm{out}\}$ denote the module under investigation and the direction of the radiation \ac{w.r.t.} the signaling molecules, respectively. In particular, $w=\mathrm{in}$ and $w=\mathrm{out}$ indicate that the molecule is illuminated and emits light, respectively.
Furthermore, due to the use of light with constant radiation wavelength $\lambda_{w}^{m}$, each photon $\Phi_{w}^{m}$ has the same energy $E_{w}^{m}$ \cite[Eq.~(1.2)]{balzani2014photochemistry}
\scaleAlign
  \begin{align}
    E_{w}^{m} = h f_{w}^{m} = h \frac{c}{\lambda_{w}^{m}}\;,
    \label{eq:E_photon}
\end{align}
where $h$, $c$, and $f_{w}^{m}$ denote the Planck's constant, the speed of light, and the radiation frequency, respectively.

The general photochemical reaction between a molecule in state $X$ with $X \in \{\mathrm{A}, \mathrm{B}\}$ and one photon $\gls{PhotonInAbs}$ is defined as follows \cite[Eqs.~(1.3),~(12.18),~(12.19)]{balzani2014photochemistry}
\scaleAlign
\begin{equation}
\begin{tikzcd}[row sep=tiny]
& & {Y} \\[-0.4cm]
& {X} + \gls{PhotonInAbs} \arrow[ur, "k_{X \rightarrow Y}" near end] \arrow[dr, "\overline{k}_{{X \rightarrow Y}}" xshift=-0.4cm] & \qquad \;, \\[-0.4cm]
& & {X}
\end{tikzcd}
\label{general_process}
\end{equation}
where $k_{X \rightarrow Y}$ and $\overline{k}_{{X \rightarrow Y}}$ denote the reaction rate constants for yielding a molecule in state $Y$ and state $X$ as reaction product, respectively. Here, all reaction rate constants have the unit $\si{\per \second}$. In \Figure{block_diagram_TX_EX}, the corresponding photochemical reactions at \ac{EX} and \ac{TX} with $(X, Y, k_{X \rightarrow Y}, \overline{k}_{{X \rightarrow Y}})$ $\underset{\scriptscriptstyle{m = \mathrm{EX}}}{\hat{=}} (\mathrm{A}, \mathrm{B}, k_{\mathrm{A}\rightarrow\mathrm{B}}, \overline{k}_{\mathrm{A}\rightarrow\mathrm{B}})$ and $(X, Y, k_{X \rightarrow Y}, \overline{k}_{{X \rightarrow Y}}) \underset{\scriptscriptstyle{m = \mathrm{TX}}}{\hat{=}} (\mathrm{B}, \mathrm{A}, k_{\mathrm{B}\rightarrow\mathrm{A}}, \overline{k}_{\mathrm{B}\rightarrow\mathrm{A}})$, respectively, are visualized.
An important measure of the efficiency of the photochemical reaction in \Equation{general_process} is the reaction quantum yield $\varphi_{{X}}$, which assumes values between $0$ and $1$, i.e., $\varphi_{{X}} = [0, 1]$, and is defined as the fraction of the reaction rate constants \cite[Eqs.~(12.18),~(12.20)]{balzani2014photochemistry}
\scaleAlign
\begin{align}
  \varphi_{{X}} = \frac{k_{X \rightarrow Y}}{k_{X \rightarrow Y} + \overline{k}_{{X \rightarrow Y}}}\;.
    \label{quantumYield}
\end{align}
For a constant total number of photons in the system, definition \Equation{quantumYield} is equivalent to the ratio of the change of the concentration of the molecules in state $X$, $\frac{\mathrm{d}C_{{X}}(t)}{\mathrm{d}t}$, in unit time to the change of the concentration of the absorbed photons, $\frac{\mathrm{d}C_{\gls{PhotonInAbs}}(t)}{\mathrm{d}t}$, in unit time, i.e.,
\scaleAlign
\begin{align}
  \varphi_{{X}} = -\frac{\mathrm{d}C_{{X}}(t)}{\mathrm{d}t} \bigg/ \frac{\mathrm{d}C_{\gls{PhotonInAbs}}(t)}{\mathrm{d}t} \label{quantumYield2}\;.
\end{align}

Here, the concentration of absorbed photons, the concentration of molecules in state $X$, and the derivative \ac{w.r.t.} time are denoted as $C_{\gls{PhotonInAbs}}(t)$, $C_{{X}}(t)$ and $\frac{\mathrm{d}}{\mathrm{d}t}$, respectively. In the extreme case, for $\varphi_{{X}} = 1$, each absorbed photon successfully triggers the molecule switching from state $X$ to state $Y$.

Not all photons, which are emitted into the fluid medium by the irradiation light, are absorbed. However, only absorbed photons contribute to the photochemical process, of course. Hence, the photon absorption process is detailed next, which characterizes the ratio of the flux of the emitted photon concentration $\frac{\mathrm{d}C_{\gls{PhotonInZero}}(t)}{\mathrm{d}t} = \frac{q_{\mathrm{in}}^{m}}{V^{m} N_{\mathrm{Av}}} \overset{\mathrm{Eq.} \Equation{eq:Light_source_to_photons}}{=}  \frac{P_{\mathrm{in}}^{m} }{ N_{\mathrm{Av}} H E_{\mathrm{in}}^{m}}$, to the flux of the absorbed photon concentration $\frac{\mathrm{d}C_{\gls{PhotonInAbs}}(t)}{\mathrm{d}t}$, both with unit $\si{\mathrm{mol} \, \metre^{-3} \second^{-1}}$. In the following, all concentration fluxes have the unit $\si{\mathrm{mol} \, \metre^{-3} \second^{-1}}$. Here, $C_{\gls{PhotonInZero}}(t)$ denotes the concentration of the photons emitted by module $m$. The absorption of a photon initiates the photochemical reaction, and is governed by the Beer-Lambert law \cite[Eqs.~(12.21),~(12.22)]{balzani2014photochemistry}
\begin{align}
  \frac{\mathrm{d}C_{\gls{PhotonInAbs}}(t)}{\mathrm{d}t} &= \frac{P_{\mathrm{in}}^{m}}{  E_{\mathrm{in}}^{m} H  N_{\mathrm{Av}}} \mkern-4mu \bigg(\mkern-4mu 1-\exp\mkern-4mu \bigg(\mkern-4mu- \log(10) \epsilon^{m} \, H C_{{X}}(t)\mkern-6mu\bigg)\mkern-6mu\bigg)\mkern-6mu\nonumber\\
  &\!\!\!\!\overset{{\mathrm{Eq.} \Equation{quantumYield2}}}{=} - \frac{1}{\varphi_{{X}}} \frac{\mathrm{d}C_{{X}}(t)}{\mathrm{d}t} \;,
  \label{eq:beer_concentration}
\end{align}
where $\epsilon^{m}$ denotes the molar absorption coefficient of the signaling molecules in $\si{\meter^2 \mathrm{mol}^{-1}}$ for the wavelength used in module $m$.
Note that \Equation{eq:beer_concentration} is independent of $A^{m}$, and the duct height $H$ is the maximum absolute distance a photon can propagate through the fluid medium.

\begin{prop}
Given an initial concentration of the molecules in state $X$ of $C_{{X}}(t=0) = C_{{X}, 0}$ at the beginning of the radiation at $t=0$, the concentration of molecules in state $X$ as a function of the irradiation time is obtained from \Equation{eq:beer_concentration} as follows
\scaleAlign
\begin{align}
  C_{{X}}(t) &= \frac{1}{a} \log\bigg[1-\exp\left(-\varphi_{{X}} \,a \,  \frac{P_{\mathrm{in}}^{m}}{  E_{\mathrm{in}}^{m} H  N_{\mathrm{Av}}} \, t \right)\nonumber \\
  & \qquad \times \, \Big(1-\exp\big(a \, C_{{X}, 0} \big)\Big)\bigg] ,
  \label{diff_solution}
\end{align}
where $a = \log(10) H \epsilon^{m}$.
\end{prop}
\begin{IEEEproof}
  Please find the proof in \Appendix{section:appendix:switching}.
\end{IEEEproof}

\scaleSubsection
\subsection{Derivation of Switching and Detection Probabilities}\label{section_photochemic_at_the_sections}
\scaleSubsectionBelow

In the following, we individually model the photochemical processes in each section of the pipe channel, i.e., \ac{EX}, \ac{TX}, propagation channel, and \ac{RX}. We refer to \Figures{block_diagram_TX_EX}{block_diagram_RX} for an overview of the possible state conversions.

\scaleSubsubsection
\subsubsection{Photochemical Reaction at EX}\label{photoChemicEX}
\scaleSubsubsectionBelow
The molecule state switching process at the \ac{EX} is given by the following chemical reaction, cf. \Figure{block_diagram_TX_EX},
\scaleAlign
\begin{align}
  \mathrm{A } +  \gls{PhotonInEX}\, \overset{k_{\mathrm{A}\,\rightarrow\,\mathrm{B}}}{\longrightarrow} \, \mathrm{B}\label{OFF_switching_process},
\end{align}
i.e., the \ac{EX} impacts only state A molecules.
As described in \Section{EX_details}, the \ac{EX} constantly radiates light into the subjacent volume $\gls{VolumeEX}$.

Let $C_{\mathrm{A},0}^{\scriptscriptstyle{\mathrm{EX}}}$ denote the concentration of molecules in state A at $z = \gls{zaEX}$, i.e., at the left boundary of the \ac{EX}. $C_{\mathrm{A},0}^{\scriptscriptstyle{\mathrm{EX}}}$ is random and further specified in \Section{ssSec:InEX}.
The concentration of molecules in state A leaving the \ac{EX} volume $\gls{VolumeEX}$, $C_{\mathrm{A}}^{\scriptscriptstyle{\mathrm{Sys}}}(\gls{TEX})$, i.e., entering volume $\mathrm{S}$ at the right boundary of \ac{EX} at $z = \gls{zbEX}$, is given by \Equation{diff_solution} with $m = \mathrm{EX}$, $X = \mathrm{A}$, $Y = \mathrm{B}$, and $C_{{X}, 0} = C_{\mathrm{A},0}^{\scriptscriptstyle{\mathrm{EX}}}$. The probability that a given molecule in state A is switched to state B during the time $\gls{TEX}$ it spends in $\gls{VolumeEX}$ is given \mbox{with \Assumption{assumptionEX} by}
\scaleAlign
\begin{align}
  \gls{prSwitchEX} = 1 - \frac{C_{\mathrm{A}}^{\scriptscriptstyle{\mathrm{Sys}}}(\gls{TEX})}{C_{\mathrm{A},0}^{\scriptscriptstyle{\mathrm{EX}}}} \;.
  \label{eq:p_s_EX}
\end{align}
\begin{remark}
Due to \ac{EX} radiation, the probability of a signaling molecule to be in state A decreases as it propagates through \ac{EX}. Therefore, the probability \ac{w.r.t.} the position, where a signaling molecule switches its state within \ac{EX}, is non-constant and decreases with increasing $z$ in \ac{EX}. However, randomness of the switching position within \ac{EX} does not impact the molecule propagation, as the characteristics of the molecules in state A and state B are similar, see \Section{propagation}. Hence, we model the switching position due to irradiation at \ac{EX} to be at $\gls{zbEX}$, i.e., the right boundary of \ac{EX}. 
\end{remark}

\scaleSubsubsection
\subsubsection{Photochemical Reaction at TX}\label{photoChemicTX}
\scaleSubsubsectionBelow

The modulation process at the \ac{TX} is described by the following chemical reaction, cf. \Figure{block_diagram_TX_EX},
\scaleAlign
\begin{align}
  \mathrm{B } +   \gls{PhotonInTX}\, \overset{k_{\mathrm{B}\,\rightarrow\,\mathrm{A}}}{\longrightarrow} \, \mathrm{A}\label{ON_switching_process},
\end{align}
i.e., the \ac{TX} impacts only state B molecules.
In contrast to the \ac{EX}, which permanently irradiates light, the \ac{TX} irradiates for a short duration of length $\gls{TTX}$ if a binary $1$ is sent.

Let $C_{\mathrm{B}}^{\scriptscriptstyle{\mathrm{TX}}}(t=0) = C_{\mathrm{B},0}^{\scriptscriptstyle{\mathrm{TX}}}$ denote the concentration of the molecules in state B in $\gls{VolumeTX}$ at the beginning of information transmission at $t=0$. $C_{\mathrm{B}}^{\scriptscriptstyle{\mathrm{TX}}}(t=0)$ is random and further specified in \Section{ssSec:InTX}. The concentration of the molecules in state B in $\gls{VolumeTX}$ as a function of time $C_{\mathrm{B}}^{\scriptscriptstyle{\mathrm{TX}}}(t)$ follows from \Equation{diff_solution} with $m = \mathrm{TX}$, $X = \mathrm{B}$, $Y = \mathrm{A}$, $C_{{X}, 0} = C_{\mathrm{B},0}^{\scriptscriptstyle{\mathrm{TX}}}$, and \Assumption{cond_static_TX}. Then, the probability of any molecule in state B in $\gls{VolumeTX}$ to be switched from state B to state A within the irradiation time $\gls{TTX}$ follows as
\scaleAlign
\begin{align}
  \gls{prSwitchTX} = 1 - \frac{C_{\mathrm{B}}^{\scriptscriptstyle{\mathrm{TX}}}(\gls{TTX})}{C_{\mathrm{B},0}^{\scriptscriptstyle{\mathrm{TX}}}} \;.
  \label{eq:pswitch}
\end{align}

\scaleSubsubsection
\subsubsection{Spontaneous Switching}\label{spSwitching}
\scaleSubsubsectionBelow
Next to the desired switching processes in \ac{EX} and \ac{TX}, photochromic molecules in state B can spontaneously switch to state A without any external trigger, while spontaneous switching from state A to state B is not possible \cite{lacombat2017photoinduced}. We model this spontaneous switching with rate constant $k_{\mathrm{s}}$ as a first-order reaction, i.e., B $\,\xrightarrow{k_{\mathrm{s}}}\,$ A. Hence, the probability $p_{\mathrm{s,SP}}$ that a state B molecule has spontaneously switched in time interval $T^{\scriptscriptstyle{\mathrm{Ch}}}(z_{\alpha}, z_{\beta})$ follows as
\scaleAlign
\begin{align}
  p_{\mathrm{s,SP}}(z_{\alpha}, z_{\beta}) &= 1-\exp\left(-k_{\mathrm{s}} T^{\scriptscriptstyle{\mathrm{Ch}}}(z_{\alpha}, z_{\beta})\right) \nonumber \\
  &= 1-\exp\left(-\frac{T^{\scriptscriptstyle{\mathrm{Ch}}}(z_{\alpha}, z_{\beta})}{T_{1/2}} \ln(2)\right)\;,
  \label{eq:spontaneous_switching}
\end{align}
where $T^{\scriptscriptstyle{\mathrm{Ch}}}(z_{\alpha}, z_{\beta})=\frac{ z_{\beta} - z_{\alpha}}{v}$ and $T_{1/2} = \frac{\ln(2)}{k_{\mathrm{s}}}$ denote the average duration a molecule needs to propagate from $z = z_{\alpha}$ to $z = z_{\beta}$ and the half-time of molecules in state B, respectively. The half-time specifies the time needed for a reaction to halve the starting concentration. The spontaneous switching can occur anywhere. However, we model it to take place only in the propagation channel between \ac{TX} and \ac{RX}, which is an accurate approximation for small \ac{EX} and \ac{TX} lengths, respectively, cf. \Assumption{assump:spontaneous_switch}. Therefore, spontaneous switching does not impact the \ac{EX} and \ac{TX} processes, respectively.

Moreover, as the propagation properties of state A and state B molecules are identical, the exact position of the spontaneous switching within the propagation channel from \ac{TX} to \ac{RX} is insubstantial. Hence, we examine the impact of spontaneous switching only for the signaling molecules which are at \ac{RX} at the sampling time. In particular, each signaling molecule entering the \ac{RX} has spontaneously switched with a probability of $\gls{prSwitchSpon} = p_{\mathrm{s,SP}}(\gls{zbEX}, \gls{zaRX})$ during propagation from \ac{EX} to \ac{RX}.
\begin{remark}
  For achieving reliable information transmission, spontaneous switching is not desirable. In contrast to molecule degradation, spontaneous molecule switching perturbs the communication link as it creates random noise conveying misleading information.
\end{remark}

\scaleSubsubsection
\subsubsection{Fluorescence at the RX}\label{fluorRX}
\scaleSubsubsectionBelow
At the \ac{RX}, molecules in state A fluoresce in response to irradiation by light of wavelength $\gls{lambdaRXin}$. In particular, at the \ac{RX}, molecules in state A can assume two substates, cf. \Figure{block_diagram_RX}. The photons emitted by the \ac{RX} light switch some of the state A molecules temporarily into an excited, non-stable substate of state A, which we denote as substate $\textnormal{A}^*$. The non-excited substate of signaling molecules in state A is denoted as ground state $\textnormal{A}^{\mathrm{G}}$ for completeness. In the absence of \ac{RX} irradiation, all state A molecules are in substate $\textnormal{A}^{\mathrm{G}}$.

The concentration of molecules in state A in \gls{VolumeRX}, $C_{\mathrm{A}}^{\scriptscriptstyle{\mathrm{RX}}}$, is constant during the fluorescence process of duration \gls{TRX} according to \Assumption{RX_static}, i.e.,
\begin{align}
  C_{\mathrm{A}}^{\scriptscriptstyle{\mathrm{RX}}} = C_{\mathrm{A}^{\mathrm{G}}}^{\scriptscriptstyle{\mathrm{RX}}}(t) + C_{\mathrm{A}^{*}}^{\scriptscriptstyle{\mathrm{RX}}}(t), \quad \mathrm{for} \quad t_\mathrm{s} \leq t \leq t_\mathrm{s} + \gls{TRX}\;.
  \label{N_A_subversions}
\end{align}
Here, $C_{\mathrm{A}^{\mathrm{G}}}^{\scriptscriptstyle{\mathrm{RX}}}(t)$ and $C_{\mathrm{A}^{*}}^{\scriptscriptstyle{\mathrm{RX}}}(t)$ denote the concentrations of the signaling molecules in substates $\textnormal{A}^{\mathrm{G}}$ and $\textnormal{A}^{*}$, respectively. $C_{\mathrm{A}}^{\scriptscriptstyle{\mathrm{RX}}}$ is constant during the fluorescence process, but can be different for every transmission, i.e., $C_{\mathrm{A}}^{\scriptscriptstyle{\mathrm{RX}}}$ is random and further specified in \Section{ssSec:RX}.

We utilize two time intervals at the \ac{RX}. Using two time intervals helps to distinguish the photons emitted from the \ac{RX} from the photons observed by the \ac{RX}. In particular, the difference between the employed wavelengths at the \ac{RX}, denoted as Stokes shift, i.e., $\gls{lambdaRXout} - \gls{lambdaRXin}$, is small for \ac{GFPD}, leading to interference issues in practical systems, if they occur simultaneously. In the proposed \ac{RX} model, in the first time interval of duration $T_\mathrm{eq}$ the \ac{RX} irradiates light and excites some of the state A molecules. In the second time interval of duration $T_{\mathrm{obs}}$, the excited molecules are counted. Hence, $\gls{TRX} = T_\mathrm{eq} + T_{\mathrm{obs}}$ follows. Next, the photochemical processes in the individual intervals are detailed.

At the beginning of the read out process at the \ac{RX}, all state A molecules are in ground substate $\textnormal{A}^{\mathrm{G}}$, i.e., $C_{\mathrm{A}^{G}}^{\scriptscriptstyle{\mathrm{RX}}}(t_\mathrm{s}) = C_{\mathrm{A}}^{\scriptscriptstyle{\mathrm{RX}}}$. At $t = t_\mathrm{s}$, the \ac{RX} starts to illuminate $\gls{VolumeRX}$ for a duration of $T_\mathrm{eq}$. The irradiation excites molecules in substate $\textnormal{A}^{\mathrm{G}}$, i.e., the \ac{RX} induces the switching from $\textnormal{A}^{\mathrm{G}}$ to $\textnormal{A}^*$. Subsequently, a molecule in substate $\textnormal{A}^*$ can switch back to substate $\textnormal{A}^{\mathrm{G}}$ and may emit light. In particular, with rate constant $k_{\mathrm{r}}$ a photon of wavelength $\gls{lambdaRXout}$ is radiated and with rate constant $k_{\mathrm{nr}}$ the switching between the substates produces heat instead of a photon, see \Figure{block_diagram_RX}. Therefore, the aforementioned process can be modeled as follows
\scaleAlign
\begin{align}
  \frac{\mathrm{d}C_{\mathrm{A}^{\mathrm{G}}}^{\scriptscriptstyle{\mathrm{RX}}}(t)}{\mathrm{d}t} &= (k_{\mathrm{r}} + k_{\mathrm{nr}}) C_{\mathrm{A}^*}^{\scriptscriptstyle{\mathrm{RX}}} - \frac{\mathrm{d}C_{\gls{PhotonInRX}}(t)}{\mathrm{d}t},
  \label{fluorescence}
\end{align}
for $t_\mathrm{s} \leq t \leq t_\mathrm{s} + T_\mathrm{eq}$, where the absorbed photon concentration flux $\frac{\mathrm{d}C_{\gls{PhotonInRX}}(t)}{\mathrm{d}t}$ is obtained from \Equation{eq:beer_concentration} as follows
\scaleAlign
\begin{align}
  \frac{\mathrm{d}C_{\gls{PhotonInRX}}(t)}{\mathrm{d}t} &\mkern-3mu=\mkern-3mu \frac{\gls{PowerRXin}}{  \gls{EnergyPhotonRX} H  N_{\mathrm{Av}}} \mkern-3mu \bigg(\mkern-4mu 1\mkern-3mu-\mkern-3mu\exp\mkern-4mu\bigg(\mkern-6mu- \log(10) \gls{epsilonRX} \, H C_{\mathrm{A}^{\mathrm{G}}}^{\scriptscriptstyle{\mathrm{RX}}}(t)\mkern-3mu\bigg)\mkern-7mu\bigg)\mkern-3mu \nonumber \\
  &\mkern-3mu=\mkern-3mu \mathrm{const.} \;
 \label{beer_lambert_rule_fluorescence}
\end{align}
\begin{remark}
  We note that during $T_\mathrm{eq}$, a molecule can change its substate multiple times from the ground substate to the excited substate and back.
\end{remark}

The duration of irradiation $T_\mathrm{eq}$ is chosen such that the state A substates reach an equilibrium, where on average the number of molecules switched from substate $\textnormal{A}^{\mathrm{G}}$ to substate $\textnormal{A}^*$ equals the number of molecules switched from substate $\textnormal{A}^*$ to substate $\textnormal{A}^{\mathrm{G}}$ in one unit time step. This can be assured for $T_\mathrm{eq} \geq T_\mathrm{eq,min}$, where $T_\mathrm{eq,min}$ denotes the minimum irradiation duration needed at the \ac{RX} to reach the equilibrium. We refer to \Appendix{Equilibrium_state_time} for the derivation of $T_\mathrm{eq,min}$. According to \Equation{eq:Teq_min}, $T_\mathrm{eq,min}$ is inversely proportional to $k_{\mathrm{r}} + k_{\mathrm{nr}}$, i.e., the equilibrium is reached faster for larger rate constants. In equilibrium, the concentration of the molecules in substate $\textnormal{A}^{\mathrm{G}}$ \mbox{in $\gls{VolumeRX}$, $C_{\mathrm{A}^{\mathrm{G}}}^{\scriptscriptstyle{\mathrm{RX}}}$, is constant}
\begin{align}
  \frac{\mathrm{d}C_{\mathrm{A}^{\mathrm{G}}}^{\scriptscriptstyle{\mathrm{RX}}}(t)}{\mathrm{d}t} = 0, \quad \mathrm{for} \quad t_\mathrm{s} + T_\mathrm{eq,min} \leq t \leq t_\mathrm{s} + T_\mathrm{eq} \;.
  \label{equilibrium_fluorescence}
\end{align}

\begin{prop}
The equilibrium concentration of state A molecules in excited substate $\mathrm{A}^*$ is obtained from \EquationsList{fluorescence}{equilibrium_fluorescence} as follows
\scaleAlign
\begin{align}
  C_{\mathrm{A}^*,\,\mathrm{eq}}^{\scriptscriptstyle{\mathrm{RX}}} = \frac{1}{x_1} - \frac{1}{x_0} W\left(\frac{x_0 x_2 \exp\left(\frac{x_0}{x_1}\right)}{x_1}\right) \;,
  \label{excited_equilibrium_number}
\end{align}
where $x_0 \mkern-3mu= \mkern-3mu\log(10) \gls{epsilonRX} H$, $x_1 \mkern-3mu =\mkern-3mu  \frac{ \gls{EnergyPhotonRX} H N_{\mathrm{Av}} (k_{\mathrm{r}} + k_{\mathrm{nr}})}{\gls{PowerRXin}}$, and $x_2 \mkern-3mu=\mkern-3mu \exp\left(-x_0 C_{\mathrm{A}}^{\scriptscriptstyle{\mathrm{RX}}} \right)$, respectively.
\end{prop}
\begin{IEEEproof}
  Please find the proof in \Appendix{section:appendix:fluorescence}.
\end{IEEEproof}
Thus, the fraction $\gls{prAStern}$ of molecules in substate $\mathrm{A}^*$ at equilibrium is given by
\scaleAlign
\begin{align}
  \gls{prAStern} =   \frac{C_{\mathrm{A}^*,\,\mathrm{eq}}^{\scriptscriptstyle{\mathrm{RX}}}}{C_{\mathrm{A}}^{\scriptscriptstyle{\mathrm{RX}}}} \;.
  \label{eq:excitedProbability}
\end{align}
After the equilibrium is reached, at $t_\mathrm{obs} = t_\mathrm{s} + T_\mathrm{eq}$ the \ac{RX} irradiation stops. Hereupon, the number of photons $N_{\gls{PhotonOutRX}}$ caused by the switching from substate $\mathrm{A}^*$ to substate $\mathrm{A}^{\mathrm{G}}$ are counted by the \ac{RX} for a time interval of $T_{\mathrm{obs}}$. We assume $T_{\mathrm{obs}} \gg \frac{1}{k_\mathrm{r}}$ to ensure all molecules in substate $\mathrm{A}^*$ have enough time to switch back to the ground substate $\mathrm{A}^{\mathrm{G}}$. Hence, $N_{\gls{PhotonOutRX}}$ can be obtained as follows
\scaleAlign
\begin{align}
  N_{\gls{PhotonOutRX}} \overset{T_{\mathrm{obs}} \gg \frac{1}{k_\mathrm{r}}}{=} \underbrace{\frac{k_{\mathrm{r}}}{k_{\mathrm{r}} + k_{\mathrm{nr}}}}_{\varphi_{\mathrm{F}}} \frac{C_{\mathrm{A}^*,\,\mathrm{eq}}^{\scriptscriptstyle{\mathrm{RX}}}}{ \gls{VolumeSizeRX} N_{\mathrm{Av}} } \;. \label{Eq:ObservationRX}
\end{align}
Finally, given a molecule in excited substate $\mathrm{A}^*$ at $t_\mathrm{obs}$, we count that molecule indirectly by observing the photon, which is emitted with probability $\gls{prPhi}$ as the molecule switches back from $\mathrm{A}^*$ to $\mathrm{A}^{\mathrm{G}}$. Hence, the probability of a substate $\mathrm{A}^*$ molecule to be detected by the \mbox{receiver is given as}
\scaleAlign
\begin{align}
  \gls{prPhi} = \varphi_\mathrm{F} = \frac{k_{\mathrm{r}}}{k_{\mathrm{r}} + k_{\mathrm{nr}}} \;.
  \label{excitation_of_photon}
\end{align}
In summary, a given state A molecule in \gls{VolumeRX} is detected with probability $\gls{prDet} = \gls{prAStern}\gls{prPhi}$ \mbox{during $\gls{TRX}$.}

\scaleSection
\section{Analytical End-to End Channel Model}
\label{sec:math}
\scaleSectionBelow
In this section, we provide insight into the propagation of the molecules. In particular, we derive the probability of arrival at the \ac{RX} for molecules which were within the \ac{TX} and outside the \ac{TX} region during modulation, respectively. In addition, the arrival probability for $s=0$, where we do not need to differentiate between inside and outside the \ac{TX}, is given. These probabilities are referred to as \ac{CIRs}\acused{CIR}. As the propagation characteristics of state A and state B molecules, respectively, are identical, the same applies to their \ac{CIRs}. Hence, in \Section{math_sec:cir}, for derivation of the \ac{CIR}, the state of the signaling molecule is irrelevant. Moreover, in \Section{ssSec:IM}, we develop a statistical model for the number of photons detected by the \ac{RX} by interpreting the transmission process of the proposed \ac{MC} system as a multistage stochastic process.

\scaleSubsection
\subsection{Channel Impulse Responses}\label{math_sec:cir}
\scaleSubsectionBelow

Here, we derive the probability of arrival at the \ac{RX} for a molecule with given position $z$ during modulation.
Due to the assumptions of uniform flow and reflective boundaries, and the considered \ac{RX} model, the molecule propagation in the system proposed in \Section{general_system_model} can be equivalently modeled as a molecule propagation in an \ac{1D} domain with infinite extent, i.e., $-\infty < z < \infty$. We derive the probability that a molecule is inside the \ac{RX} given that it was inside the \ac{TX} at position $z^{\scriptscriptstyle{\mathrm{TX}}}$ at time $t \in [0, 0 + \gls{TTX}]$, as a function of time, i.e., the irradiation starts at $t=0$ and stops at $t = \gls{TTX}$. We refer to this probability as $h_0(t)$. In the following, we approximate $t \in [0, 0 + \gls{TTX}]$ by $t=0$, cf. \Assumption{assump_modulation_time} in \Appendix{sections_assumptions}.
The position $z^{\scriptscriptstyle{\mathrm{TX}}}$ of the signaling molecules is random and uniformly distributed \mbox{in the \ac{TX} region $[\gls{zaTX}, \gls{zbTX}]$.}
\begin{prop}
The probability $h_0(t)$ that a signaling molecule, whose position is random and uniformly distributed in \gls{VolumeTX}, i.e., $\gls{zaTX} \leq z^{\scriptscriptstyle{\mathrm{TX}}} \leq \gls{zbTX}$, at time $t=0$, is inside the \ac{RX} region, i.e., $\gls{zaRX} \leq z^{\scriptscriptstyle{\mathrm{RX}}} \leq \gls{zbRX}$, at time $t$ is given by
\scaleAlign
\begin{align}
  h_0(t) &= \frac{1}{2 \,\gls{lengthTX}} \sum_{i=0}^3 (-1)^{i} \left[  a_i \erf\mkern-4.5mu\left(\frac{a_i}{\sqrt{ 4 D \, t}}\right) \right.  \nonumber \\
  &\qquad \left. +  \sqrt{\frac{4 D \, t}{\pi}}  \exp\left( \frac{- a_i^2}{4 D \, t}\right)   \right] \;,
   \label{cir_analytic}
\end{align}
where $\{ a_0, a_1, a_2, a_3\} = \{ \gls{zbRX} - \gls{zaTX} - v t, \gls{zbRX} - \gls{zbTX} - v t, \gls{zaRX} - \gls{zbTX} - v t, \gls{zaRX} - \gls{zaTX} - v t \}$ and $\erf(x)$ denotes the Gaussian error function.
\end{prop}

\begin{IEEEproof}
  Please find the proof in \Appendix{section:appendix:IR}.
\end{IEEEproof}

Similarly, we obtain the probability $h_1(t)$ that a signaling molecule, which was in $\gls{VolumeTXNot} = \mathrm{S} \setminus \gls{VolumeTX}$ at $t=0$, i.e., not in the \ac{TX} region during \ac{TX} irradiation, is inside the \ac{RX} region as follows
\scaleAlign
\begin{align}
h_1(t) &=\frac{1}{2 \left(\gls{lengthSys} - \gls{lengthTX}\right)} \sum_{i=0}^3 (-1)^{i} \left[b_i \erf\left(\frac{b_i}{\sqrt{ 4 D \, t}}\right)  \right.\nonumber \\
& \qquad \left. +  \sqrt{\frac{4 D \, t}{\pi}}  \exp\left( \frac{- b_i^2}{4 D \, t}\right)   \right]  \;,
\label{cir_h1_analytic}
\end{align}
where $\{ b_0, b_1, b_2, b_3\} = \{ \gls{zbRX} - \gls{zbTX} - v t, \gls{zbRX} - \gls{zbSys} - v t, \gls{zaRX} - \gls{zbSys} - v t, \gls{zaRX} - \gls{zbTX} - v t \}$.

Furthermore, the probability $h_2(t)$ that a signaling molecule is inside the \ac{RX} region, if it was in $\gls{VolumeTX} \cup \gls{VolumeTXNot} = \mathrm{S}$ at $t=0$, follows as
\scaleAlign
\begin{align}
  h_2(t) &= \frac{1}{2 \left(\gls{lengthSys}\right)} \sum_{i=0}^3 (-1)^{i} \left[ c_i \erf\mkern-4.5mu\left(\frac{c_i}{\sqrt{ 4 D \, t}}\right)  \right. \nonumber \\
  & \qquad \left. +  \sqrt{\frac{4 D \, t}{\pi}}  \exp\left( \frac{- c_i^2}{4 D \, t}\right)   \right] \;,
   \label{cir_h2_analytic}
\end{align}
where $\{ c_0, c_1, c_2, c_3\} = \{ \gls{zbRX} - \gls{zaTX} - v t, \gls{zbRX} - \gls{zbSys} - v t, \gls{zaRX} - \gls{zbSys} - v t, \gls{zaRX} - \gls{zaTX} - v t \}$.
Note that $h_2(t)$ is related to $h_0(t)$ and $h_1(t)$ as $h_2(t) = h_0(t) \frac{\gls{lengthTX}}{\gls{lengthSys}} + h_1(t) \frac{\gls{lengthSys}-\gls{lengthTX}}{\gls{lengthSys}}$.

\scaleSubsection
\subsection{Statistical Model}\label{ssSec:IM}
\scaleSubsectionBelow

\begin{figure}[!tbp]
  \centering
  \includegraphics[width=1\columnwidth]{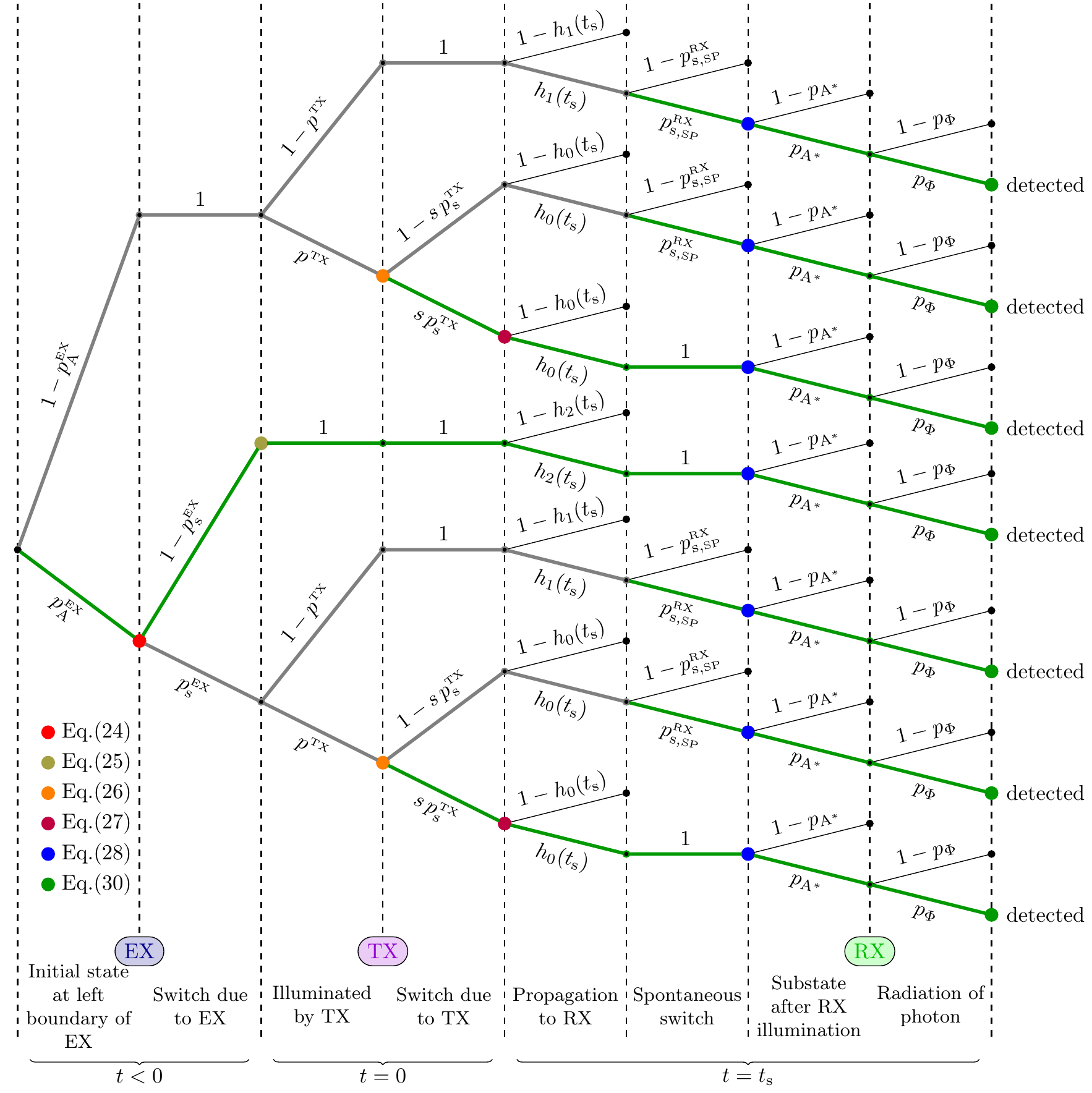}
  \caption{The probability tree of the conglomerate of eight stochastic processes is shown. Each level represents one of the consecutive stochastic processes indicated by the dashed lines with corresponding description. The levels are consecutive \ac{w.r.t.} time and we extend only the paths which result in an observation of a photon $\gls{PhotonOutRX}$. A branch with label $1$ shows that the current level does not impact the molecule. The colored branches indicate the state of the molecule, where green and gray represent state A and state B, respectively. The colored nodes correspond to \EquationsList{into_EX}{nrMolRX}, \Equation{nrPhotonsRX}.}
  \label{probability_trees}
\end{figure}

In this section, we determine the statistics of the number of photons detected by the \ac{RX}. These photons originate from the fluorescence reaction of state A molecules at the \ac{RX} (see \Section{fluorRX}). \Figure{probability_trees} shows that there are several possibilities for a signaling molecule to be detected by the \ac{RX} based on the photon $\gls{PhotonOutRX}$ it emits. In particular, the fate of any individual molecule is characterized by a conglomerate of eight consecutive stochastic processes, where the output of one process is the input of the subsequent process. Each random process has two possible outcomes and therefore can be modeled as a Bernoulli process. This process is repeated for every input molecule. Eventually, each outcome occurrence follows a Binomial distribution, as the input molecules are modeled to be mutually statistically independent. The latter claim is proved in this section. Finally, we determine the  probability distribution of the number of photons detected at the \ac{RX} characterized by the last stochastic process in the proposed system, which is conditional on all previous processes.

According to \Section{general_system_model}, $\gls{NSys}$ signaling molecules are uniformly distributed in S at time $t=0$. Therefore, each of these signaling molecules has traversed the \ac{EX} during $t<0$. At $t=0$, the \ac{TX} radiates light if $s=1$, and the \ac{RX} counts the signaling molecules in $\gls{VolumeRX}$ at $t=t_\mathrm{s}$ for subsequent detection, cf. \Figure{probability_trees}.

\scaleSubsubsection
\subsubsection{State Distribution of Molecules entering \ac{EX}}\label{ssSec:InEX}
\scaleSubsubsectionBelow
All $\gls{NSys}$ signaling molecules have entered \ac{EX} at $t<0$, but only a subset $N_{\mathrm{A}}^{\scriptscriptstyle{\mathrm{EX}}} \leq \gls{NSys}$ entered \ac{EX} in state A. In particular, a given signaling molecule entered \ac{EX} in state A with a probability of $ \gls{prAEX}$, and with a probability of $1-\gls{prAEX}$ in state B.
Hence, the state in which a molecule enters \ac{EX} can be modeled as a Bernoulli random variable. Therefore, $N_{\mathrm{A}}^{\scriptscriptstyle{\mathrm{EX}}}$ follows the Binomial distribution
\scaleAlign
\begin{align}
  N_{\mathrm{A}}^{\scriptscriptstyle{\mathrm{EX}}} \sim \binomial{\gls{NSys}, \gls{prAEX}} \;.
  \label{into_EX}
\end{align}

\scaleSubsubsection
\subsubsection{Initial Number of State A Molecules}\label{ssSec:OutEX}
\scaleSubsubsectionBelow
When a state A molecule propagates through \ac{EX}, it is switched to a state B molecule with probability $\gls{prSwitchEX}$, see \Equation{eq:p_s_EX}. From \Equations{diff_solution}{eq:p_s_EX}, we observe that $\gls{prSwitchEX}$ depends on the number of nearby state A molecules, i.e., the concentration $C_{\mathrm{A},0}$ of signaling molecules in state A. In particular, for large $C_{\mathrm{A},0}$, $\gls{prSwitchEX}$ is influenced by the competition of molecules for photons. Therefore, during the switching process, there is a statistical dependence between the states of different molecules and the Binomial distribution does not apply. However, the competition can be neglected for small $C_{\mathrm{A},0}$. In this case, $\gls{prSwitchEX}$ is constant \ac{w.r.t.} $C_{\mathrm{A},0}$, which is what we assume in the following. The validity of this assumption is verified in \Section{sec:switching_process_verification}.
Hence, applying $\textit{Theorem 1}$ on conditional Binomial distributions in \Appendix{section:appendix:conditionalProbabilities}, the overall number of state A molecules in S at $t=0$ follows the Binomial distribution
\scaleAlign
\begin{align}
 N_{\mathrm{A}}^{\scriptscriptstyle{\mathrm{Sys}}} \mkern-5mu \sim \binomial{\mkern-4mu \gls{NSys}, \gls{prAEX} \mkern2mu (1-\gls{prSwitchEX})\mkern-4mu}\mkern2mu.
 \label{into_system}
\end{align}

\scaleSubsubsection
\subsubsection{Number of Molecules at \ac{TX}}\label{ssSec:InTX}
\scaleSubsubsectionBelow
$N_{\mathrm{B}}^{\scriptscriptstyle{\mathrm{Sys}}} = \gls{NSys} - N_{\mathrm{A}}^{\scriptscriptstyle{\mathrm{Sys}}}$ state B molecules are uniformly distributed in S at $t=0$. Hence, $N_{\mathrm{B}}^{\scriptscriptstyle{\mathrm{Sys}}} \sim \binomial{\gls{NSys}, (1-\gls{prAEX}) + \gls{prAEX} \gls{prSwitchEX}}$ follows also a Binomial distribution, as $N_{\mathrm{B}}^{\scriptscriptstyle{\mathrm{Sys}}}$ and $N_{\mathrm{A}}^{\scriptscriptstyle{\mathrm{Sys}}}$ are fully correlated. However, only a subset of the these molecules $N_{\mathrm{B}}^{\scriptscriptstyle{\mathrm{TX}}} \leq N_{\mathrm{B}}^{\scriptscriptstyle{\mathrm{Sys}}}$, is in $\gls{VolumeTX}$ at $t=0$. In particular, with a probability of $ \gls{prTX} =  \frac{\gls{VolumeSizeTX}}{\gls{VolumeSizeSys}}$ a given signaling molecule M is inside the \ac{TX} volume, i.e., $\mathrm{M} \in \gls{VolumeTX}$, and with a probability of $1-\gls{prTX}$ it is outside the \ac{TX}. The availability of a molecule for information transmission can be modeled as a Bernoulli random variable.
Therefore, the total number of state B molecules $N_{\mathrm{B}}^{\scriptscriptstyle{\mathrm{TX}}}$ available at the \ac{TX} follows the Binomial distribution
\scaleAlign
\begin{align}
  N_{\mathrm{B}}^{\scriptscriptstyle{\mathrm{TX}}}  &\sim \binomial{N_{\mathrm{B}}^{\scriptscriptstyle{\mathrm{Sys}}}, \gls{prTX}} \nonumber \\
  &\!\!\!\!\overset{\mathrm{Eq.}\Equation{eq:condBinom}}{=} \binomial{\gls{NSys}, (1-\gls{prAEX}) \gls{prTX} + \gls{prAEX} \gls{prSwitchEX}\gls{prTX}}\;.
  \label{TX_B_molecules}
\end{align}

\scaleSubsubsection
\subsubsection{Number of Switched Molecules at \ac{TX}}\label{ssSec:Switched}
\scaleSubsubsectionBelow
 When a state B molecule is in $\gls{VolumeTX}$ at $t=0$, it is switched to a state A molecule with probability $s \mkern2mu \gls{prSwitchTX}$ within the modulation interval, i.e., with probability $\gls{prSwitchTX}>0$ and $\gls{prSwitchTX}=0$ for transmit symbol $s=1$ and $s=0$, respectively. From \Equation{diff_solution}, we observe that $\gls{prSwitchTX}$ depends on the initial number of molecules $N_{\mathrm{B}}^{\scriptscriptstyle{\mathrm{TX}}}$ in $\gls{VolumeTX}$. However, using the arguments from \Section{ssSec:OutEX}, we can assume the conversion probability is independent of $N_{\mathrm{B}}^{\scriptscriptstyle{\mathrm{TX}}}$. Hence, the number of state A molecules at the \ac{TX} switched from state B during modulation follows the Binomial distribution
 \scaleAlign
 \begin{align}
  N_{\mathrm{B}\rightarrow\mathrm{A}}^{\scriptscriptstyle{\mathrm{TX}}} \mkern-5mu &\sim \binomial{\mkern-4mu N_{\mathrm{B}}^{\scriptscriptstyle{\mathrm{TX}}}, s \mkern2mu \gls{prSwitchTX}} \nonumber \\
  & \!\!\!\!\overset{\mathrm{Eq.}\Equation{eq:condBinom}}{=} \binomial{\gls{NSys}, s\mkern2mu \gls{prSwitchTX}\gls{prTX} \left((1-\gls{prAEX})  +  \gls{prAEX} \gls{prSwitchEX}\right)} \;.
\end{align}
In the following, we refer to the difference between the realization, $N_{\mathrm{B}\rightarrow\mathrm{A}}^{\scriptscriptstyle{\mathrm{TX}}}$, and the expectation $\mathbb{E}\{N_{\mathrm{B}\rightarrow\mathrm{A}}^{\scriptscriptstyle{\mathrm{TX}}}\} = \gls{NSys} s\mkern2mu \gls{prSwitchTX}\gls{prTX} \left((1-\gls{prAEX})  +  \gls{prAEX} \gls{prSwitchEX}\right)$, as \ac{TX} noise $n^{\scriptscriptstyle{\mathrm{TX}}}$, i.e., $n^{\scriptscriptstyle{\mathrm{TX}}}= N_{\mathrm{B}\rightarrow\mathrm{A}}^{\scriptscriptstyle{\mathrm{TX}}} - \mathbb{E}\{N_{\mathrm{B}\rightarrow\mathrm{A}}^{\scriptscriptstyle{\mathrm{TX}}}\}$. The \ac{TX} noise is a key characteristic of media modulation based \ac{MC}. In the proposed system, the number of signaling molecules used for information transmission cannot be fully controlled as the molecules are not released by the \ac{TX}. Hence, $n^{\scriptscriptstyle{\mathrm{TX}}}$ is affected by the randomness of the state of the signaling molecule at the left boundary of the \ac{EX}, the randomness of the state conversion process in the \ac{EX}, the randomness of the availability of the signaling molecules at the \ac{TX}, and the randomness of the state conversion process in the \ac{TX}.

\scaleSubsubsection
\subsubsection{Number of State A Molecules at \ac{RX} at Sampling Time}\label{ssSec:RX}
\scaleSubsubsectionBelow
The number of signaling molecules at \ac{RX} at time $t_{\mathrm{s}}$, $N^{\scriptscriptstyle{\mathrm{RX}}}$, depends on the molecule propagation in the channel. As all molecules are assumed to propagate independently, the arrival of a molecule at \ac{RX} can be modeled as a Bernoulli random variable with success probability $h_j(t)$ according to \EquationsList{cir_analytic}{cir_h2_analytic}. Here, $j$ is chosen according to the position of a molecule at time $t=0$. Hence, $j=0$, $j=1$, and $j=2$ is selected for molecules, which are uniformly distributed in $\gls{VolumeTX}$, in $\gls{VolumeTXNot}$, and in $\mathrm{S}$, respectively.
The number of state A molecules at the \ac{RX} at sampling time $N_{\mathrm{A}}^{\scriptscriptstyle{\mathrm{RX}}}$ with $N_{\mathrm{A}}^{\scriptscriptstyle{\mathrm{RX}}} \leq N^{\scriptscriptstyle{\mathrm{RX}}}$ is also affected by the spontaneous switching, which can occur during the propagation of a state B molecule to the \ac{RX}, cf. \Section{spSwitching}. In particular, with probability $\gls{prSwitchSpon}$ a signaling molecule arrives at the \ac{RX} in state A instead of state B due to the spontaneous switching.
Thus, $N_{\mathrm{A}}^{\scriptscriptstyle{\mathrm{RX}}}$ follows the Binomial distribution
\scaleAlign
\begin{align}
  N_{\mathrm{A}}^{\scriptscriptstyle{\mathrm{RX}}}(t = t_{\mathrm{s}})  \sim  \binomial{\gls{NSys},p_{\mathrm{A}}^{\scriptscriptstyle{\mathrm{RX}}}}\;,
  \label{nrMolRX}
\end{align}
where
\scaleAlign
\begin{align}
  p_{\mathrm{A}}^{\scriptscriptstyle{\mathrm{RX}}} &= (1-\gls{prAEX})\Big[(1-\gls{prTX}) \gls{prSwitchSpon} h_1(t_{\mathrm{s}}) \nonumber \\
  & \quad \quad + \gls{prTX} \big[(1-s \gls{prSwitchTX})\gls{prSwitchSpon}h_0(t_{\mathrm{s}}) + s \gls{prSwitchTX} h_0(t_\mathrm{s}) \big]\Big] \nonumber \\
  & \quad + \gls{prAEX} \gls{prSwitchEX} \Big[ (1-\gls{prTX}) \gls{prSwitchSpon} h_1(t_\mathrm{s}) \nonumber \\
  & \quad \quad + \gls{prTX}\big[(1-\gls{prTX}) \gls{prSwitchSpon} h_0(t_\mathrm{s}) + s \gls{prSwitchTX} h_0(t_\mathrm{s})\big]\Big] \nonumber \\
  & \quad + \gls{prAEX}(1-\gls{prSwitchEX}) h_2(t_\mathrm{s}) \;,
\end{align}
cf. \Figure{probability_trees}.
\scaleSubsubsection
\subsubsection{Number of Received Photons at the \ac{RX}}\label{ssSec:RX_signal}
\scaleSubsubsectionBelow
Finally, the number of received photons, $N_{\gls{PhotonOutRX}}$, at the \ac{RX} during detection additionally depends on the fluorescence process. Each of the $N_{\mathrm{A}}^{\scriptscriptstyle{\mathrm{RX}}}$ state A molecules in the \ac{RX} can independently switch from the ground state $\mathrm{A}^{\mathrm{G}}$ to the excited substate $\mathrm{A}^{\mathrm{*}}$ and back during \ac{RX} irradiation, which starts at $t_{\mathrm{s}}$. The irradiation stops at $t_{\mathrm{obs}}$, and with a probability of $\gls{prAStern}$ the current substate of a signaling molecule is $\mathrm{A}^{\mathrm{*}}$. After $t_{\mathrm{obs}}$, an excited molecule switches back to the non-excited substate $\mathrm{A}^{\mathrm{G}}$ and emits a photon $\gls{PhotonOutRX}$ with probability $\gls{prPhi}$. Hence, the detection of a state A molecule at \ac{RX} via $\gls{PhotonOutRX}$ can be modeled as a Bernoulli random variable with success probability $\gls{prDet} = \gls{prAStern} \gls{prPhi}$ according to \Equations{eq:excitedProbability}{excitation_of_photon}. Hence, $N_{\gls{PhotonOutRX}}$ follows the Binomial distribution
\scaleAlign
\begin{align}
  N_{\gls{PhotonOutRX}}  \sim  \binomial{N_{\mathrm{A}}^{\scriptscriptstyle{\mathrm{RX}}},\gls{prDet}} \overset{\mathrm{Eq.}\Equation{eq:condBinom}}{=} \binomial{\gls{NSys},p_{\mathrm{A}}^{\scriptscriptstyle{\mathrm{RX}}} \gls{prDet}}\;.
  \label{nrPhotonsRX}
\end{align}
For simplicity of notation, we will substitute $r = N_{\gls{PhotonOutRX}}$ in the following. Finally, the probability $p_{\mathrm{r}, s} = p_{\mathrm{A}}^{\scriptscriptstyle{\mathrm{RX}}} \gls{prDet}$ denotes the probability that a signaling molecule is detected at the \ac{RX}, i.e., $p_{\mathrm{r}, s=1}$ and $p_{\mathrm{r}, s=0}$, when $1$ and $0$ was transmitted, respectively.

\scaleSubsubsection
\subsubsection{Noise Sources}\label{ssSec:Noise}
\scaleSubsubsectionBelow

In the proposed system, several sources of noise impair the information transmission. In particular, there exists background noise due to incomplete signal molecule switching at the \ac{EX} unit, \ac{TX} noise due to the initial distribution of the signaling molecules, modulation noise due to the switching process at the \ac{TX}, signal dependent background noise due to spontaneous switching in the channel, diffusion-noise caused during molecule propagation, and reception noise caused by the fluorescence based readout at the \ac{RX}. The statistical model of the received signal is given by the Binomial distribution in \Equation{nrPhotonsRX}, i.e., all considered noise sources are included in \Equation{nrPhotonsRX}. We denote the difference between the received signal and the expectation of the received signal as the effective noise, i.e., $n_{\mathrm{eff}} = r - \mathbb{E}\{r\}$, which has zero mean and variance $\sigma^2_{n_{\mathrm{eff}}} =\gls{NSys} p_{\mathrm{r}, s=1} (1-p_{\mathrm{r}, s=1})$.

\scaleSection
\section{Symbol Detection and Performance Analysis}
\label{sec:performance}
\scaleSectionBelow
In this section, we derive a threshold detection scheme based on the statistical model developed in \Section{ssSec:IM} and analyse the \ac{BER} for the proposed media modulation based \ac{MC} system.
\scaleSubsection
\subsection{Optimal Detector}\label{math_sec:ssDet}
\scaleSubsectionBelow
We apply the \ac{ML} decision rule to obtain an estimate $\hat{s}$ of the transmit symbol $s$ as follows \cite{Jamali2019ChannelMF}
\scaleAlign
\begin{align}
  \hat{s} &= \underset{s \in \{0,1\}}{\text{argmax}} \; \fpdf[r]{r \given s} \nonumber \\
    &=\mkern-4.5mu \begin{cases}
        \mkern-4.5mu 1 , \; \text{if} \; \frac{\binom{\gls{NSys}}{r} p_{\mathrm{r}, s=1}^{r} (1-p_{\mathrm{r}, s=1})^{\gls{NSys}-r}}{\binom{\gls{NSys}}{r} p_{\mathrm{r}, s=0}^{r} (1-p_{\mathrm{r}, s=0})^{\gls{NSys}-r}} \\
        \quad \quad \quad \mkern-4.5mu= \left(\frac{1-p_{\mathrm{r}, s=1}}{ 1-p_{\mathrm{r}, s=0}}\right)^{\gls{NSys}}\mkern-6.5mu\left(\frac{(1-p_{\mathrm{r}, s=0})p_{\mathrm{r}, s=1} }{ (1-p_{\mathrm{r}, s=1}) p_{\mathrm{r}, s=0}}\right)^{r}\mkern-5.5mu\geq 1 \\
        \mkern-4.5mu0 , \;\text{otherwise}
    \end{cases},
  \label{eq:math_section_ML}
\end{align}

where $\fpdf[r]{r \given s}$ denotes the probability of observing $r$ photons given symbol $s$ was transmitted, which is a Binomial distribution according to \Equation{nrPhotonsRX}.

In the following, we show that the \ac{ML} decision rule in \Equation{eq:math_section_ML} is equivalent to a threshold detector employing a single threshold value without any performance loss. The threshold value needed for detection can be computed offline and can then be utilized throughout the communication process. Therefore, the threshold detector reduces the computational \mbox{complexity for online data detection.}

\begin{prop}
  The \ac{ML} decision rule in \Equation{eq:math_section_ML} can be equivalently realized by a threshold detector employing a single decision threshold as follows
  \scaleAlign
  \begin{align}
    \hat{s} &=
        \begin{cases}
          1 , \; \text{if} \; r \geq  \xi \\
          0 , \;\text{otherwise} \;
        \end{cases}\;,
        \label{eq:math_section:threshold}
  \end{align}
  with threshold $\xi = \lceil\Theta\rceil $ and $\Theta \in \mathbb{R}^{+}_{0}$ given by
  \scaleAlign
  \begin{align}
    \Theta =
    \begin{cases}
     -\gls{NSys} \frac{\log\left(\frac{1-p_{\mathrm{r}, s=1}}{1-p_{\mathrm{r}, s=0}}\right)}{\log\left(\frac{(1-p_{\mathrm{r}, s=0})p_{\mathrm{r}, s=1} }{ (1-p_{\mathrm{r}, s=1}) p_{\mathrm{r}, s=0}}\right)} \; &, \text{if} \; p_{\mathrm{r}, s=0} >  0 \\
      \quad 1 \;&,\text{if} \; p_{\mathrm{r}, s=0} =  0 \;
    \end{cases}\;.
    \label{theta_threshold}
  \end{align}
\end{prop}
\begin{IEEEproof}
  Due to space constraints, we only provide a sketch of the proof. An optimal threshold value exists if for one unique value $\Theta \in \mathbb{R}^{+}_{0}$ with $\xi = \lceil\Theta\rceil $ the decision metric given by the \ac{ML} decision rule in \Equation{eq:math_section_ML} equals one, i.e., $\fpdf[r]{\Theta \given s=1} = \fpdf[r]{\Theta \given s=0}$. The uniqueness can be proven by showing that \Equation{eq:math_section_ML} is monotonically increasing in $r$. The value of $\Theta $ is obtained by solving $\fpdf[r]{\Theta \given s=1} = \fpdf[r]{\Theta \given s=0}$. This completes the proof.
\end{IEEEproof}

\scaleSubsection
\subsection{Bit Error Rate}\label{Bit_error_rate_derivation}
\scaleSubsectionBelow

The \ac{BER} of the proposed \ac{MC} system can be expressed as follows
\scaleAlign
\begin{align}
  &P_{\mathrm{e}} \nonumber \\
  &= \prob{s = 0} \prob{\hat{s} = 1 \given s = 0} + \prob{s = 1} \prob{\hat{s} = 0 \given s = 1} \nonumber \\
  &\overset{(a)}{=} \frac{1}{2} \sum_{r = \xi}^{\gls{NSys}} \binom{\gls{NSys}}{r} p_{\mathrm{r}, s=0}^r (1-p_{\mathrm{r}, s=0})^{\gls{NSys}-r} \nonumber \\
  &\qquad  + \frac{1}{2} \sum_{r = 0}^{\xi-1} \binom{\gls{NSys}}{r} p_{\mathrm{r}, s=1}^r (1-p_{\mathrm{r}, s=1})^{\gls{NSys}-r} \nonumber \\
  &\overset{(b)}{=} \frac{1}{2} \left[1 - \sum_{r = 0}^{\xi-1} \binom{\gls{NSys}}{r} p_{\mathrm{r}, s=0}^r (1-p_{\mathrm{r}, s=0})^{\gls{NSys}-r} \right] \nonumber \\
  & \qquad + \frac{1}{2} \sum_{r = 0}^{\xi-1} \binom{\gls{NSys}}{r} p_{\mathrm{r}, s=1}^r (1-p_{\mathrm{r}, s=1})^{\gls{NSys}-r} \nonumber \\
  &\overset{(c)}{=} \frac{1}{2} \bigg(1 - I_{1-p_{\mathrm{r}, s=0}}\left(\gls{NSys}-\xi+1, \xi\right) \nonumber \\
  &\qquad + I_{1-p_{\mathrm{r}, s=1}}\left(\gls{NSys}-\xi+1, \xi\right) \bigg)\;,
   \label{BER_derivation}
\end{align}
where we exploit in $(a)$ the threshold detection rule \Equation{eq:math_section:threshold} and the fact that $r$ is an integer value, in $(b)$ the mass function property $\sum_r \fpdf{r} = 1$, and in $(c)$ the \ac{CDF} of a Binomial distribution $\sum_{k = 0}^{x} \binom{N}{k} p^k (1-p)^{N-k} = I_{1-p}(N-x, x+1)$. Here, $I_{a}(b, c)$ denotes the regularized incomplete Beta function.

\scaleSection
\section{Performance Evaluation}
\label{sec:evaluation}
\scaleSectionBelow
In this section, we first specify the parameter values of \ac{GFPD} \cite{brakemann2011reversibly}, which we adopt as a practically feasible option for photoswitchable fluorescent molecules. Then, we evaluate the statistical model in \Equation{nrPhotonsRX}. Finally, the dependence of the \ac{BER} in \Equation{BER_derivation} on the various system parameters is evaluated.
%
%
\scaleSubsection
\subsection{Choice of Parameter Values}\label{section:evaluation:PBSAndParam}
\scaleSubsectionBelow

The default values of the adopted system parameters are given in \Table{Table:Parameter}. These values are used in the following if not specified otherwise. The \ac{GFPD} specific parameter values are taken from \cite{brakemann2011reversibly, Junghans2016DiffusionGFPD, arai2018spontaneously, uno2019reversibly, ruhlandt2020absolute, lacombat2017photoinduced}. The parameter values related to the duct are chosen such that they have the same order of magnitude as those of the human cardiovascular system \cite[Chap. 14]{hall2020guyton}. Note that all parameters used in this section satisfy the conditions of the assumptions in \Appendix{sections_assumptions}.

To verify the accuracy of the analytical expression for the statistics of the received molecules, stochastic \ac{3D} \ac{PBS} was carried out.
Note that the probabilities for state switching \gls{prSwitchEX}, \gls{prSwitchTX}, \gls{prSwitchSpon}, and \gls{prAStern} are calculated based on \Equation{eq:p_s_EX}, \Equation{eq:pswitch}, \Equation{eq:spontaneous_switching}, and \Equation{eq:excitedProbability}, respectively, and are then adopted for \ac{PBS}. \ac{PBS} is employed to verify \gls{prTX}, $h_0(t_{\mathrm{s}})$, $h_1(t_{\mathrm{s}})$, $h_2(t_{\mathrm{s}})$, and the statistical model derived for the consecutive stochastic processes given in \Equation{nrPhotonsRX}\footnote{A more accurate \ac{PBS} model, which determines \gls{prSwitchEX}, \gls{prSwitchTX}, \gls{prSwitchSpon}, and \gls{prAStern} by simulating the related photochemical processes, does not seem computationally feasible. In fact, a power density $\gls{PowerTX} = 1 \times 10^{6} \, \si{\watt \per\m \squared}$ would require the simulation of approximately $9 \times 10^{5}$ photons per timestep $\Delta t = 1 \times 10^{-2} \,\si{\second}$, which does not seem possible.}.
The results from \ac{PBS} were averaged over $10^{4}$ realizations. In addition, we used Monte Carlo simulation to validate the analytical expression for the \ac{BER} in \Equation{BER_derivation}. Hereby, we randomly generated $10^{6}$ transmit symbols. Next, according to the transmit symbols, values for $N_{\gls{PhotonOutRX}}$ were randomly generated based on the proposed and validated statistical model. Finally, the \ac{BER} for all possible threshold values up to $\xi \leq 100$ was numerically determined and the lowest \ac{BER} selected.
\begin{table*}[!tbp]
\caption{Default Values for Simulation Parameters.}
\begin{center}
  \begin{threeparttable}
{\def\arraystretch{1.5}\tabcolsep=12pt
 \begin{tabular}{|l | c | c | r|}
   \hline
 Parameter & Description & Value & Ref. \\ [0.1ex]
 \hline\hline
 $H$, $W$ & Duct height, duct width & $0.001 \,\si{\meter}$, $0.001 \,\si{\meter}$ & \cite{hall2020guyton} \\
 \hline
 $\gls{zaSys}$, $\gls{zbSys}$ & Left and right boundary of S & $0 \,\si{\meter}$, $0.5 \,\si{\meter}$ &  \\
 \hline
 $\gls{zaEX}$, $\gls{zbEX}$ & Left and right boundary of \ac{EX} & $-0.01 \,\si{\meter}$, $0 \,\si{\meter}$ &\\
 \hline
 $\gls{zaTX}$, $\gls{zbTX}$ & Left and right boundary of \ac{TX} & $0 \,\si{\meter}$, $0.05 \,\si{\meter}$ &\\
 \hline
 $\gls{zaRX}$, $\gls{zbRX}$ & Left and right boundary of \ac{RX} & $0.4 \,\si{\meter}$, $0.45 \,\si{\meter}$ &\\
 \hline
 $v$ & Flow velocity &  $0.01 \, \si{\meter \per \second}$ & \cite{hall2020guyton}\\
 \hline
 $D$ & Diffusion coefficient & $1 \times 10^{-10}\, \si{\meter \squared \per \second} $ &\cite{Junghans2016DiffusionGFPD} \\
 \hline
 $ \gls{NSys}$ & Number of molecules in S & $1000$ & \\
 \hline
 $\gls{PowerEX}$ & Irradiation power density at $\textnormal{EX}^{(\textnormal{a})}$ & $1 \times 10^{6} \, \si{\watt \per\m \squared}$ &\cite{brakemann2011reversibly}\\
 \hline
 $\gls{PowerTX}$ & Irradiation power density at $\textnormal{TX}^{(\textnormal{b})}$ & $1 \times 10^{6} \, \si{\watt \per\m \squared}$ &\cite{brakemann2011reversibly}\\
 \hline
 $\gls{PowerRXin}$ & Irradiation power density at $\textnormal{RX}^{(\textnormal{c})}$ & $1 \times 10^{10} \, \si{\watt \per\m \squared}$ &\cite{brakemann2011reversibly}\\
 \hline
 $\gls{lambdaEX}$ & Wavelength to switch GFP at \ac{EX} & $405 \times 10^{-9} \,\si{\meter}$ &\cite{brakemann2011reversibly}\\
 \hline
 $\gls{lambdaTX}$ & Wavelength to switch GFP at \ac{TX} & $365 \times 10^{-9} \,\si{\meter}$ &\cite{brakemann2011reversibly}\\
 \hline
 $\gls{lambdaRXin}$ & Wavelength to trigger fluorescence & $511 \times 10^{-9} \,\si{\meter}$ &\cite{brakemann2011reversibly}\\
 \hline
 $\gls{lambdaRXout}$ & Wavelength of fluorescence & $529 \times 10^{-9} \,\si{\meter}$ &\cite{brakemann2011reversibly}\\
 \hline
 $\gls{epsilonEX}$ & Molar absorption coefficient @$405$nm & $1.9 \times 10^{3} \,\si{\meter^2 \mathrm{mol}^{-1}}$ &\cite{arai2018spontaneously}\\
 \hline
 $\gls{epsilonTX}$ & Molar absorption coefficient @$360\mathrm{nm}^{(\textnormal{d})}$ & $2.2 \times 10^{3} \,\si{\meter^2 \mathrm{mol}^{-1}}$ &\cite{arai2018spontaneously}\\
 \hline
 $\gls{epsilonRX}$ & Molar absorption coefficient @$511$nm & $8.3 \times 10^{3} \,\si{\meter^2 \mathrm{mol}^{-1}}$ &\cite{brakemann2011reversibly}\\
 \hline
 $\varphi_{\mathrm{B}}$ & Reaction quantum yield B$\to$A & $2.4 \times 10^{-2}$ &\cite{uno2019reversibly}\\
 \hline
 $\varphi_{\mathrm{A}}$ & Reaction quantum yield A$\to$B & $3.4 \times 10^{-3}$ &\cite{uno2019reversibly}\\
 \hline
 $\varphi_{\mathrm{F}}$ & Fluorescence quantum yield &  $0.47$ &\cite{ruhlandt2020absolute}\\
 \hline
 $T_{1/2}$ & Half time in state B &  $600 \,\si{\second}$ &\cite{lacombat2017photoinduced}\\
 \hline
  $\tau_{\mathrm{A}^*} = \frac{1}{k_{\mathrm{r}}+ k_{\mathrm{nr}}}$ & Excited state lifetime &  $2.9 \times 10^{-9} \,\si{\second}$ &\cite{ruhlandt2020absolute}\\[0.05cm]
 \hline
 \rule{0pt}{12pt}$\gls{TEX}$ & Irradiation time at $\textnormal{EX}^{(\textnormal{e})}$ & $ \frac{\gls{zbEX}-\gls{zaEX}}{v} $ &\cite{brakemann2011reversibly}\\
 \hline
 $\gls{TTX}$ & Irradiation time at $\textnormal{TX}^{(\textnormal{f})}$ & $ 5 \times 10^{-3} \,\si{\second}$ &\cite{brakemann2011reversibly}\\
 \hline
 $T_{\mathrm{eq}}$ & Irradiation time at $\textnormal{RX}^{(\textnormal{g})} \,^{(\textnormal{h})}$ & $ 1 \times 10^{-6} \,\si{\second}$ &\cite{brakemann2011reversibly}\\
 \hline
 $T_{\mathrm{obs}}$ & Observation duration at $\textnormal{RX}^{(\textnormal{i})}$ & $ 1 \times 10^{-6} \,\si{\second}$ &\\
 \hline
 $\epsilon_{\mathrm{eq}}$ & Precision error w.r.t. equilibrium concentration & $  10^{-6} \,\si{\second^{-1}}$ &\\
 \hline
 $\chi$ & Condition threshold \ac{w.r.t.} assumptions in \Appendix{sections_assumptions} & $  100$ &\\
 \hline
 $\Delta t$ & Time step \ac{PBS} & $ 1 \times 10^{-2} \,\si{\second}$ &\\
 \hline
 \end{tabular}
 }
 \begin{tablenotes}[flushleft]\footnotesize\setlength\itemsep{0cm}
\item[(a)] \cite{brakemann2011reversibly} reported values between $2 \times 10^{4} \, \si{\watt \per\m \squared}$ and $4.3 \times 10^{10} \, \si{\watt \per\m \squared}$.
\item[(b)] \cite{brakemann2011reversibly} reported values between $1 \times 10^{3} \, \si{\watt \per\m \squared}$ and $1.6 \times 10^{6} \, \si{\watt \per\m \squared}$.
\item[(c)] \cite{brakemann2011reversibly} reported values between $8.2 \times 10^{3} \, \si{\watt \per\m \squared}$ and $4 \times 10^{7} \, \si{\watt \per\m \squared}$.
\item[(d)] \,No value for $365\mathrm{nm}$ given in literature. Hence, we assume molar absorption coefficient @$360\mathrm{nm} \hat{=} $@$365\mathrm{nm}$.
\item[(e)] \cite{brakemann2011reversibly} reported values between $1 \times 10^{-7} \, \si{\second}$ and $4 \times 10^{2} \, \si{\second}$.
\item[(f)] \cite{brakemann2011reversibly} reported values between $1 \times 10^{-7} \, \si{\second}$ and $4 \times 10^{2} \, \si{\second}$.
\item[(g)] \cite{brakemann2011reversibly} reported values between $1 \times 10^{-2} \, \si{\second}$ and $1.5 \times 10^{-1} \, \si{\second}$.
\item[(h)] $T_{\mathrm{eq,min}} \approx 4 \times 10^{-8} \si{\second} \;$ for default values, cf. \Equation{eq:Teq_min}. The maximum value observed for the parameter values used in our simulations was $T_{\mathrm{eq,min}} \approx 8 \times 10^{-8} \si{\second} $.\vspace{0.15cm}
\item[(i)] $\frac{1}{k_{\mathrm{r}}} =  \frac{\tau_{\mathrm{A}^*}}{\varphi_{\mathrm{F}}} = 6.18 \times 10^{-9} \si{\second} \ll T_{\mathrm{obs}}\;$, $\,$cf. \Equation{Eq:ObservationRX}.
\end{tablenotes}
\end{threeparttable}
\end{center}
 \label{Table:Parameter}
 \vspace{-0.5cm}
\end{table*}

\scaleSubsection
\subsection{Evaluation of the Switching Process and the Statistical Model} \label{sec:switching_process_verification}
\scaleSubsectionBelow

\begin{figure*}[t]
    \centering
    \begin{minipage}[t]{0.48\textwidth}
      \centering
      \includegraphics[width = \textwidth, trim={0 0 0 1.3cm},clip]{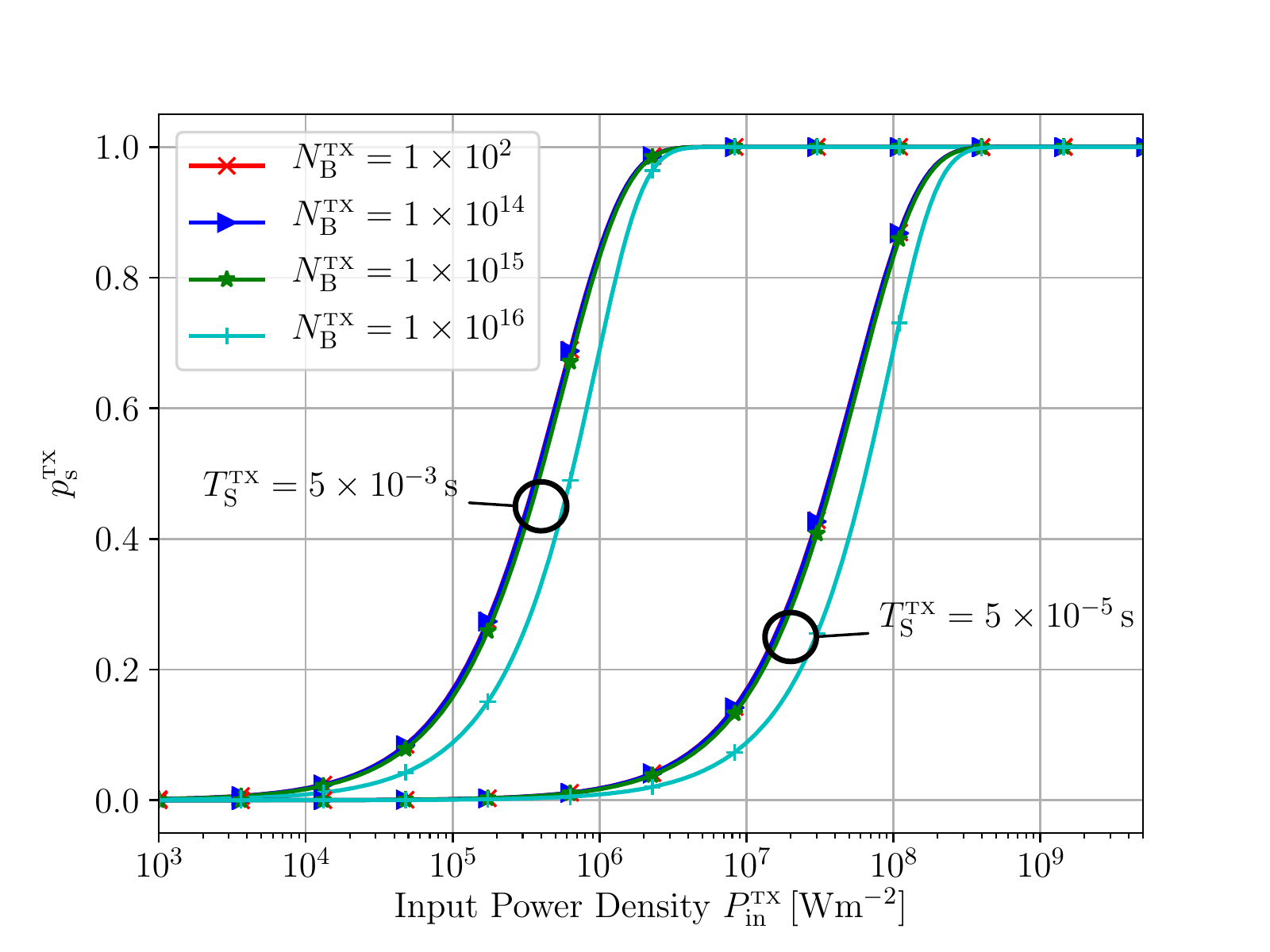}
      \caption{The probability of a molecule within the \ac{TX} to switch from state B to state A within modulation time lengths $\gls{TTX} = 5 \times 10^{-3} \,\si{\second}$ and $\gls{TTX} = 5 \times 10^{-5} \,\si{\second}$ as a function of the input power density $\gls{PowerTX}$ for different numbers of molecules $N_{\mathrm{B}}^{\scriptscriptstyle{\mathrm{TX}}}$ in $\gls{VolumeTX}$.}
      \label{fig:switching_characteristic}
    \end{minipage}
    \hfill
  \begin{minipage}[t]{0.48\textwidth}
    \centering
      \includegraphics[width = \textwidth, trim={0 0 0 1.3cm},clip]{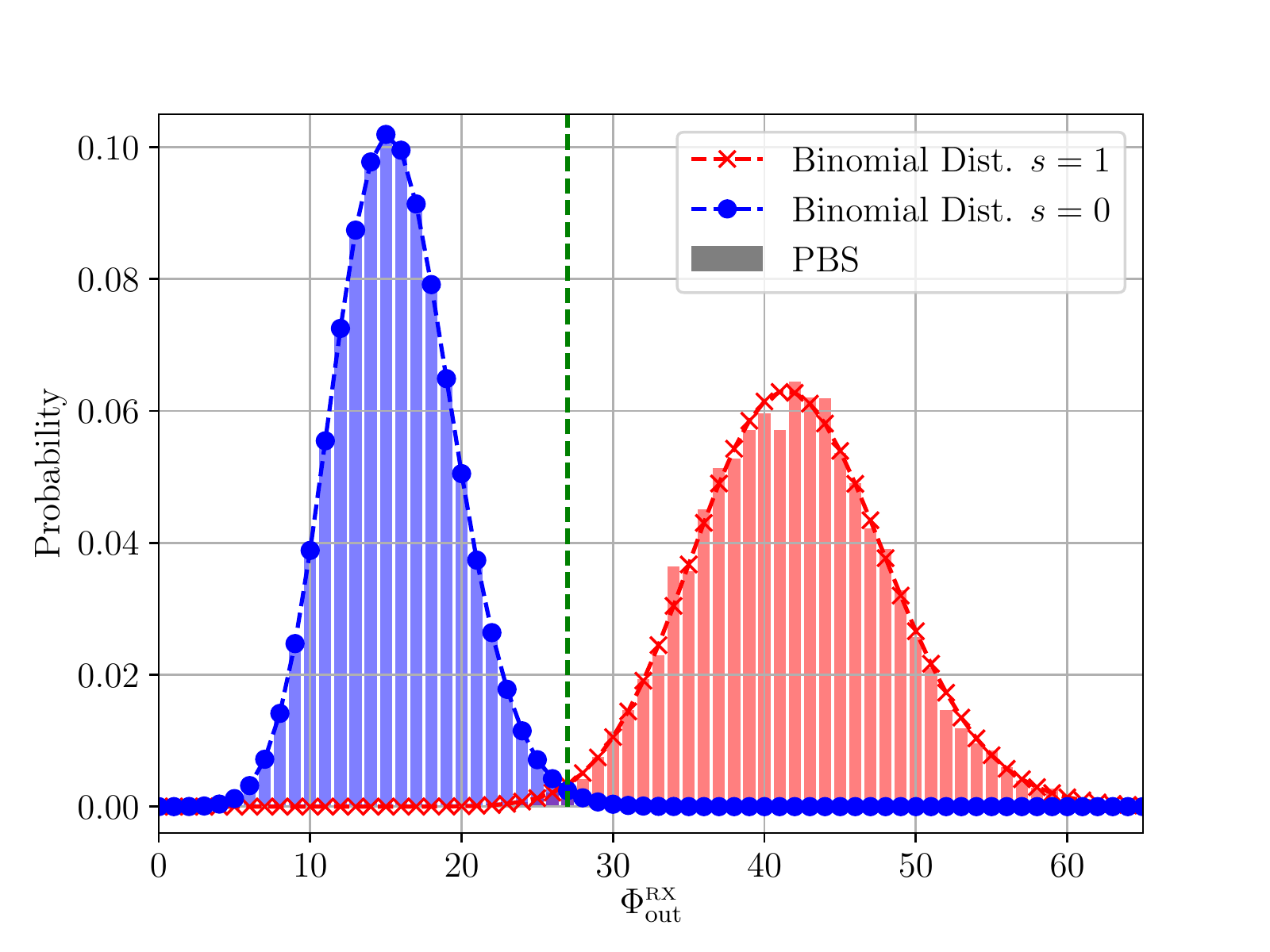}
      \caption{The empirical (PBS) and analytical distribution of the number of received photons $N_{\gls{PhotonOutRX}}$ according to \Equation{nrPhotonsRX} for $s=1$ and $s=0$ and the optimal threshold value (vertical dashed line). Here, illumination power densities $\gls{PowerEX} =  1 \times 10^{4} \,\si{\watt \per\m \squared}$, $\gls{PowerTX} =  1 \times 10^{6} \,\si{\watt \per\m \squared}$, and $\gls{PowerRXin} =  2 \times 10^{12} \,\si{\watt \per\m \squared}$ are adopted.
      }
    \label{fig:histogram}
  \end{minipage}
\end{figure*}

First, we exemplarily investigate the photochemical reaction at the \ac{TX}, discussed in \Section{photoChemicTX}, for \ac{GFPD}. In \Figure{fig:switching_characteristic}, $\gls{prSwitchTX}$, as defined in \Equation{eq:pswitch}, is shown as a function of input light power density $\gls{PowerTX}$ for irradiation durations $ \gls{TTX} = 5 \times 10^{-3} \,\si{\second}$ and $ \gls{TTX} = 5 \times 10^{-5} \,\si{\second}$, respectively.
The range of values for $\gls{PowerTX}$ considered in this paper is large compared to \cite{brakemann2011reversibly}, where power density values ranging from $ 1 \times 10^{3} \,\si{\watt \per\m \squared}$ to  $ 1.6 \times 10^{6} \, \si{\watt \per\m \squared}$ were used for light sources with wavelength $\lambda_{\mathrm{BA}}$, cf. \Table{Table:Parameter}.
In addition, the authors in \cite{brakemann2011reversibly} show the general feasibility of lasers as light sources, e.g., for imaging. Hence, very high light power density values are in principle possible. Therefore, we also consider larger power density values compared to \cite{brakemann2011reversibly} in this paper.
From \Figure{fig:switching_characteristic}, we observe that the likelihood of a molecule to switch within the irradiation time is low for small input power density, increases for increasing input power density, and converges to $1$ for large values of $\gls{PowerTX}$. Furthermore, we observe for a given input power density that a smaller irradiation time $ \gls{TTX} = 5 \times 10^{-5} \,\si{\second}$ results in a smaller conversion probability $\gls{prSwitchTX}$.
Moreover, \Figure{fig:switching_characteristic} shows that $\gls{prSwitchTX}$ remains unchanged for a wide range of $N_{\mathrm{B}}^{\scriptscriptstyle{\mathrm{TX}}}$. Only for systems with a very large number of signaling molecules, i.e., if $N_{\mathrm{B}}^{\scriptscriptstyle{\mathrm{TX}}}\geq 10^{14}$, a larger input power density is necessary to achieve a given switching probability due to the competition of signaling molecules for photons. As the system geometry is identical at \ac{EX}, \ac{TX}, and \ac{RX}, approximating $\gls{prSwitchEX}$, $\gls{prSwitchTX}$, and $\gls{prAStern}$ as independent from the number of signaling molecules $\gls{NSys}$, $N_{\mathrm{B}}^{\scriptscriptstyle{\mathrm{TX}}}$, and $N_{\mathrm{A}}^{\scriptscriptstyle{\mathrm{RX}}}$, respectively, as done in \SectionsThree{ssSec:OutEX}{ssSec:Switched}{ssSec:RX_signal} is justified if the number of signaling molecules is sufficiently small. Therefore, as $N_{\mathrm{B}}^{\scriptscriptstyle{\mathrm{TX}}}$ and $N_{\mathrm{A}}^{\scriptscriptstyle{\mathrm{RX}}}$ are upper bounded by $\gls{NSys}$, $N_{\mathrm{B}}^{\scriptscriptstyle{\mathrm{TX}}}\leq \gls{NSys} \leq 10^{14} \land N_{\mathrm{A}}^{\scriptscriptstyle{\mathrm{RX}}}\leq \gls{NSys} \leq 10^{14} $ is sufficient to ensure statistical independence of the signaling molecules.

In \Figure{fig:histogram}, we show the probability mass function of the number of received photons $N_{\gls{PhotonOutRX}}$ according to \Equation{nrPhotonsRX} for $s=1$ and $s=0$ and compare it to results from \ac{PBS}. The corresponding optimal threshold value is highlighted by the vertical dashed line. We observe that the results obtained from \ac{PBS} and \Equation{nrPhotonsRX}, respectively, match, which confirms the statistical model proposed in \Section{ssSec:IM}.
\scaleSubsection
\subsection{Evaluation of BER} \label{sec:BER_plots}
\scaleSubsectionBelow
\begin{figure*}[t]
    \centering
    \begin{minipage}[t]{0.48\textwidth}
      \centering
      \includegraphics[width = 1\textwidth, trim={0 0 0 1.3cm},clip]{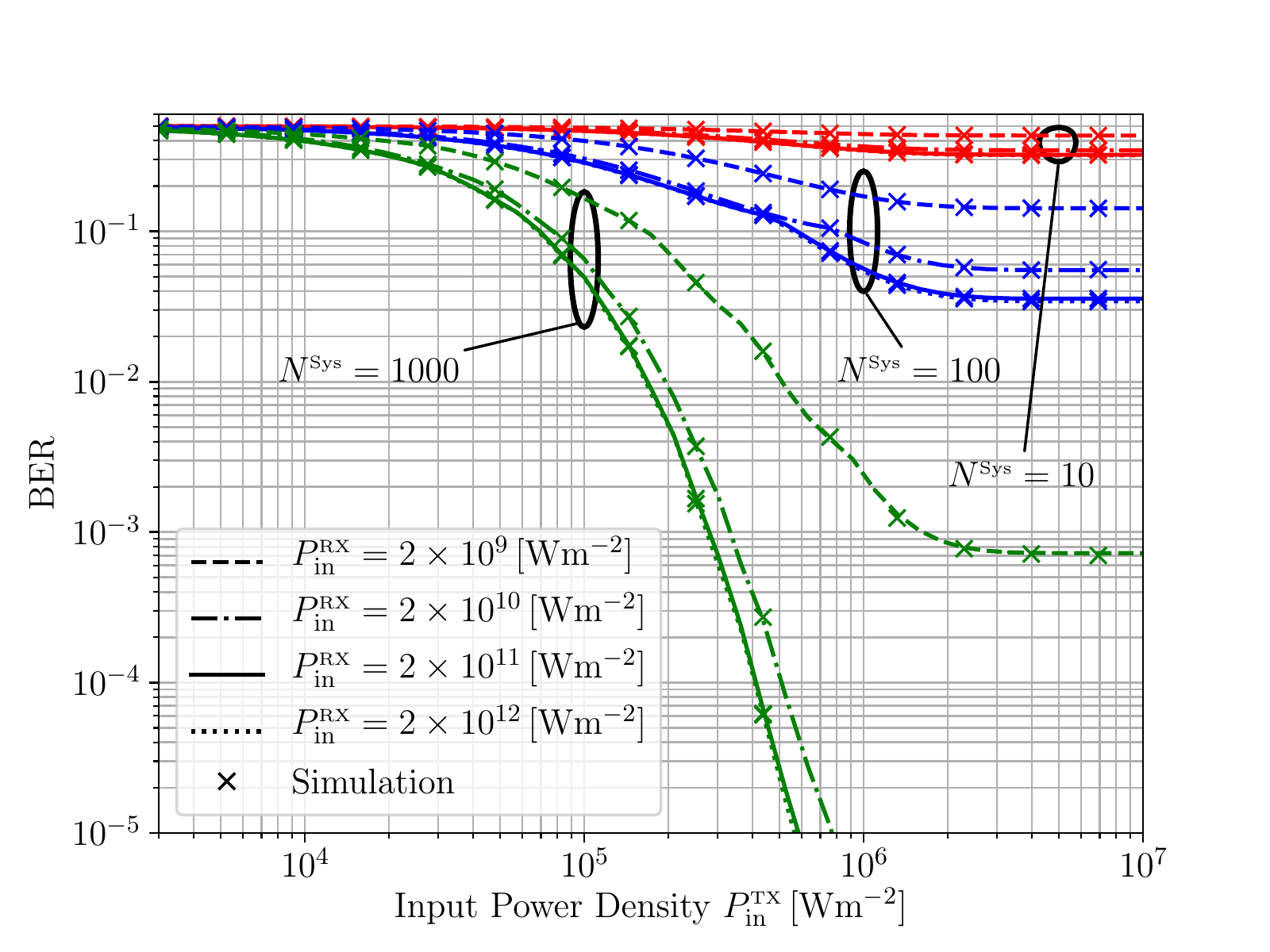}
      \caption{\ac{BER} as a function of the \ac{TX} irradiation power density for different numbers of signaling molecules and different \ac{RX} irradiation power densities. The results from Monte Carlo simulation are depicted by markers.}
      \label{fig:BER_over_TX_power}
    \end{minipage}
    \hfill
    \begin{minipage}[t]{0.48\textwidth}
      \centering
      \includegraphics[width = 1\textwidth, trim={0 0 0 1.3cm},clip]{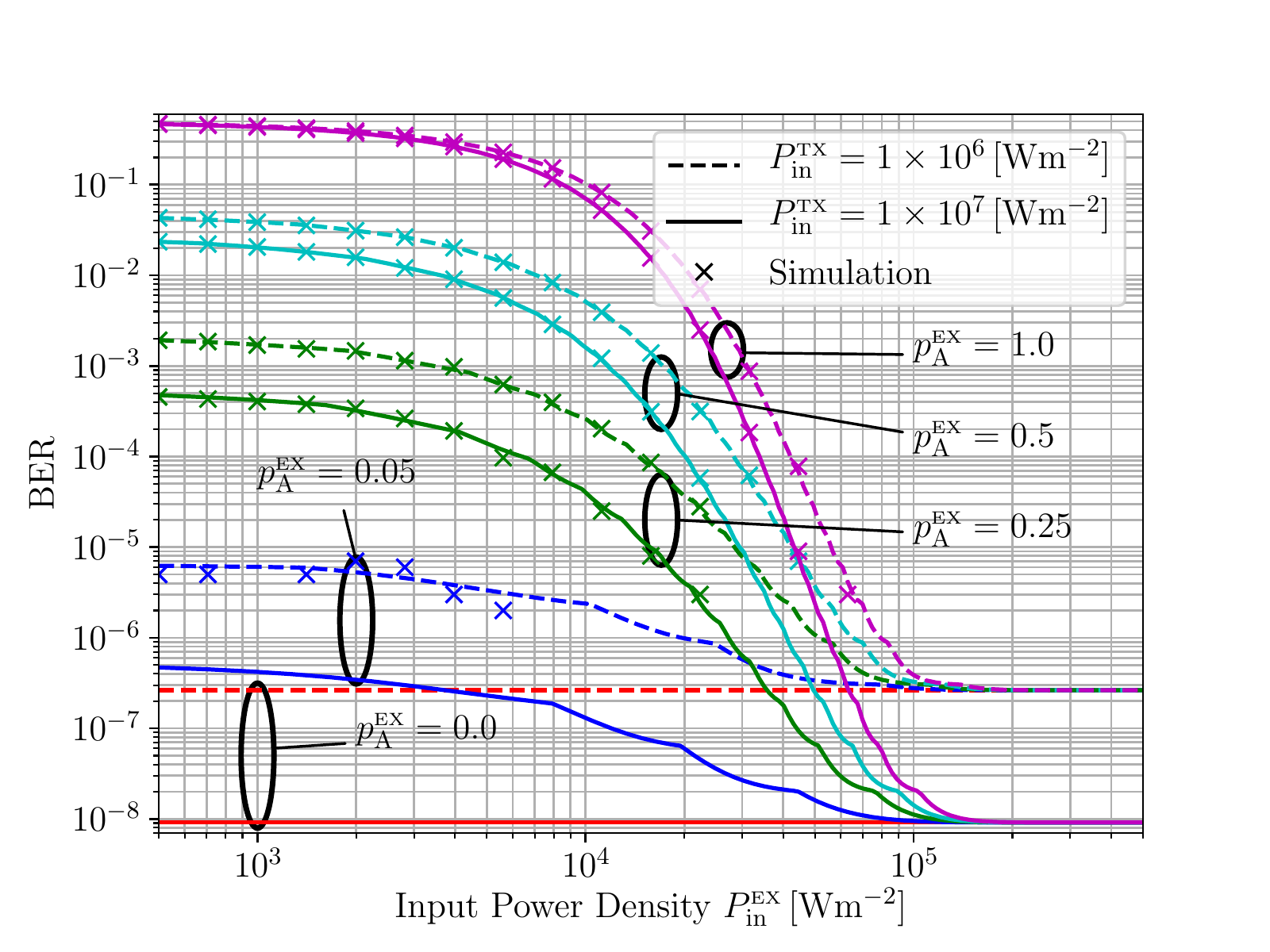}
      \caption{\ac{BER} as a function of the \ac{EX} irradiation power density for different state A probabilities $\gls{prAEX}$ of the signaling molecules at the left boundary of the \ac{EX} and different \ac{TX} irradiation power densities. The results from Monte Carlo simulation are depicted by markers.}
      \label{fig:BER_over_EX_power}
    \end{minipage}
\end{figure*}

In this section, we evaluate the \ac{BER} for the proposed system.
We mainly vary the \ac{EX}, \ac{TX}, and \ac{RX} irradiation power densities, respectively, as this can be easily realized in a practical setup. Moreover, we study the impact of the state probability $\gls{prAEX}$ of the signaling molecules at the left boundary of \ac{EX} on the \ac{BER}. Finally, we show the performance loss caused by spontaneous switching.

\scaleSubsubsection
\subsubsection{Impact of TX and RX Irradiation Power Density on BER} \label{sec:BER_power}
\scaleSubsubsectionBelow

In \Figure{fig:BER_over_TX_power}, the \ac{BER} is shown as a function of the irradiation power density $\gls{PowerTX}$ for different numbers of signaling molecules $\gls{NSys}$ and different \ac{RX} radiation power densities $\gls{PowerRXin}$. Here, $\gls{prAEX} = 0$ is assumed.
We observe that the \ac{BER} decreases as $\gls{PowerTX}$ increases. For $\gls{PowerTX} > 4 \times 10^{6} \, \si{\watt \per\m \squared}$, the \ac{BER} approaches an error floor, which is visible for $\gls{NSys} = \{10, 100\}$ for all choices of $\gls{PowerRXin}$, but occurs also for larger \gls{NSys} at lower \ac{BER} values. In particular, for large $\gls{PowerTX}$ values, $\gls{prSwitchTX} = 1$ follows, i.e., the conversion at the \ac{TX} is deterministic. The \ac{BER} value of the error floor is dependent on the transmitter noise $n^{\scriptscriptstyle{\mathrm{TX}}}$, the noise due to spontaneous switching, and the randomness in the detection process. Moreover, we observe from \Figure{fig:BER_over_TX_power} that the \ac{BER} decreases as $\gls{NSys}$ increases. This is intuitive, as the standard deviation of a Binomial distributed random variable $\sqrt{\gls{NSys} p (1-p)}$ normalized to the mean value $\gls{NSys} p$, i.e., $\frac{\sqrt{\gls{NSys} p (1-p)}}{\gls{NSys} p} = \frac{\sqrt{(1-p)}}{\sqrt{\gls{NSys} p}}$, converges to zero for $\gls{NSys} \rightarrow \infty$. Hence, for increasing numbers of signaling molecules $\gls{NSys}$, discriminating between binomial distributions $N_{\gls{PhotonOutRX}}(s=1)$ and $N_{\gls{PhotonOutRX}}(s=0)$ becomes more reliable, which results in a smaller \ac{BER}.
Next, we consider the impact of $\gls{PowerRXin}$. From \Figure{fig:BER_over_TX_power}, we observe that the BER decreases as $\gls{PowerRXin}$ increases. $\gls{PowerRXin}$ impacts the probability $\gls{prAStern}$ of a state A molecules to be excited by the \ac{RX}, which is needed for emission of a photon $\gls{PhotonOutRX}$. Moreover, we observe that the BERs for $\gls{PowerRXin} = 2 \times 10^{11} \, \si{\watt \per\m \squared}$ and $\gls{PowerRXin} = 2 \times 10^{12} \, \si{\watt \per\m \squared}$ are almost identical. In particular, for $\gls{PowerRXin} = 2 \times 10^{11} \, \si{\watt \per\m \squared}$, $\gls{prAStern} \rightarrow 1$ follows. Thus, a further increase of the \ac{RX} irradiation power density can not improve the \ac{BER} results. We finally note that the numerically and analytically determined BERs perfectly match.

\scaleSubsubsection
\subsubsection{Impact of EX Irradiation Power Density on BER} \label{sec:BER_eraser}
\scaleSubsubsectionBelow

In \Figure{fig:BER_over_EX_power}, we show the \ac{BER} as a function of the \ac{EX} irradiation power density for $\gls{prAEX} = \{1, 0.5, 0.25, 0.05, 0\}$ and two different \ac{TX} irradiation power densities. We chose $\gls{PowerRXin} = 2 \times 10^{12}\, \si{\watt \per\m \squared}$ as \ac{RX} irradiation power density to ensure $\gls{prAStern} \approx 1$, which eliminates the randomness caused by low \ac{RX} irradiation power densities.

First, we consider $\gls{PowerTX} = 1 \times 10^{6} \, \si{\watt \per\m \squared}$. We observe that for $\gls{prAEX} > 0$ the \ac{BER} decreases as $\gls{PowerEX}$ increases. In particular, an increase in $\gls{PowerEX}$ increases the number of state B signaling molecules $N_{\mathrm{B}}^{\scriptscriptstyle{\mathrm{TX}}}$, which are available for modulation at the \ac{TX}, cf. \Equation{TX_B_molecules}. Furthermore, we observe that for $\gls{prAEX} = 0.05$ and for $\gls{prAEX} = 1$ the \ac{BER} improves by a factor of $2\times 10^{1}$ and $2\times 10^{6}$, respectively, if $\gls{PowerEX}$ is increased from $\gls{PowerEX} = 5 \times 10^{2} \, \si{\watt \per\m \squared}$ to $\gls{PowerEX} = 5 \times 10^{5} \, \si{\watt \per\m \squared}$. Hence, the actual increase of $N_{\mathrm{B}}^{\scriptscriptstyle{\mathrm{TX}}}$ due to $\gls{PowerEX}$ heavily depends on $\gls{prAEX}$. From that we conclude that for large $\gls{prAEX}$, i.e., values close to one, an \ac{EX} is mandatory for the proposed system to ensure reliable communication, whereas for $\gls{prAEX}$ close to zero, the \ac{EX} can be deactivated, e.g., in case of power consumption limitations. Furthermore, for large $\gls{PowerEX}$, all signaling molecules exit \ac{EX} as state B molecules. Hence, the \ac{BER} is independent of $\gls{prAEX}$ for \mbox{large $\gls{PowerEX}$, which leads to the error floor in \Figure{fig:BER_over_EX_power}.}

Now, we consider the impact of $\gls{PowerTX}$. We observe that the \ac{BER} decreases if $\gls{PowerTX}$ is increased from $1 \times 10^{6} \, \si{\watt \per\m \squared}$ to $1 \times 10^{7} \, \si{\watt \per\m \squared}$. However, the \ac{BER} reduction depends on both $\gls{prAEX}$ and $\gls{PowerEX}$. In particular, an increase of the \ac{TX} irradiation power density has no visible impact on the \ac{BER} if $\gls{prAEX} = 1$ and $\gls{PowerEX}$ is small, whereas for $\gls{prAEX} = 0$ or large $\gls{PowerEX}$, we observe a large improvement of the \ac{BER}.

Finally, we observe that, for very small BERs, the results obtained by Monte Carlo simulation do not perfectly match the analytical results since, in this case, the number of Monte Carlo iterations was too low to obtain numerically stable results.

\scaleSubsubsection
\subsubsection{Impact of Spontaneous Switching and Flow Velocity on BER} \label{sec:BER_flow}
\scaleSubsubsectionBelow

In the proposed \ac{MC} system, the state of the signaling molecules is used to encode information. In contrast to \ac{MoSK}, where the types of molecules employed are assumed to be independent, in media modulation, there exists a dependency between the different states A and B even after the modulation process at the \ac{TX} due to the spontaneous switching from state B to state A. We show in the following that the amount of noise introduced by the spontaneous switching of \ac{GFPD} is velocity dependent and results in a performance degradation.

\begin{figure}[!tbp]
  \centering
  \includegraphics[width = 1\columnwidth, trim={0 0 0 1.3cm},clip]{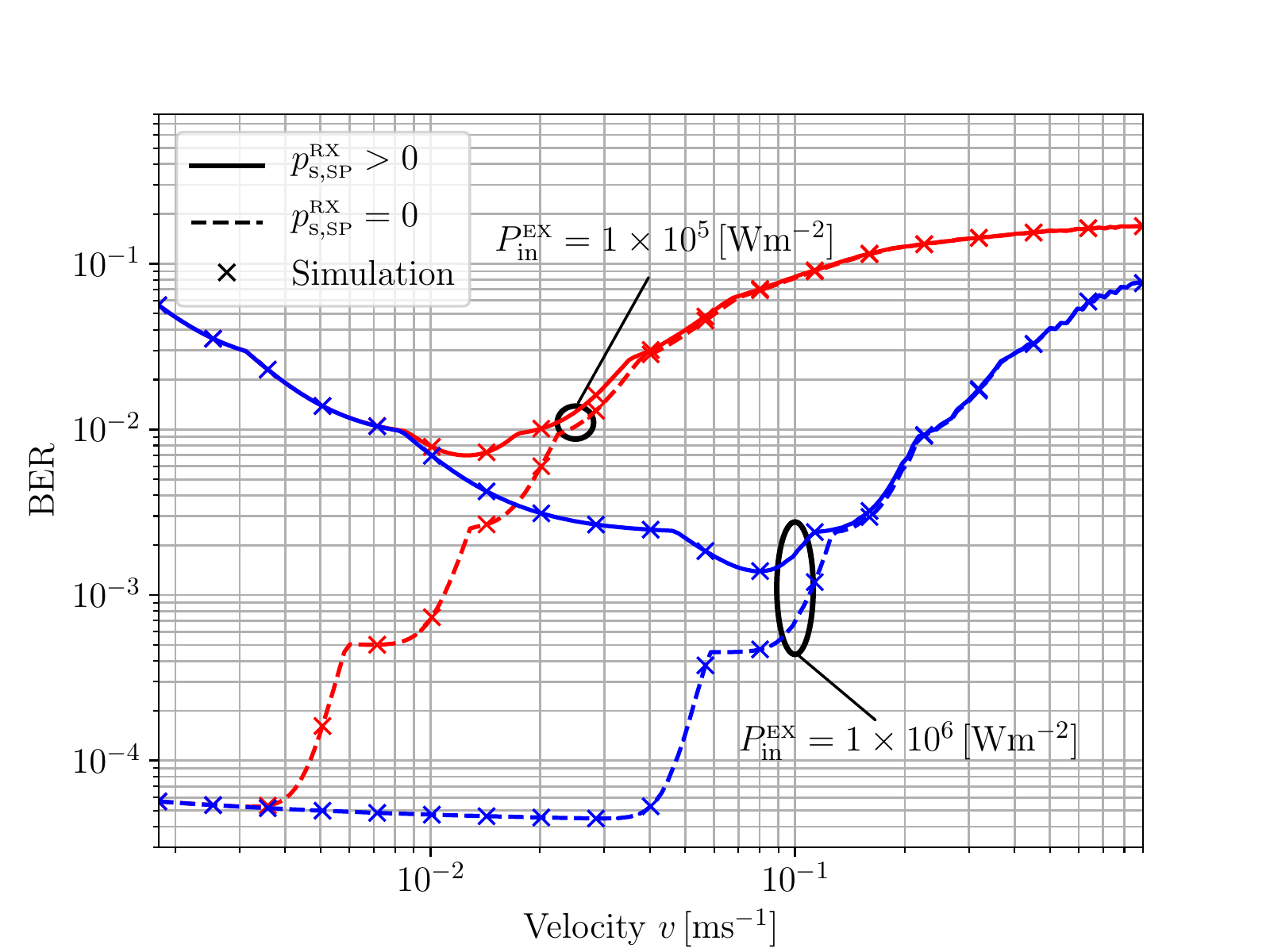}
  \caption{BER as a function of flow velocity $v$ for different \ac{EX} radiation power densities. We shown results in dashed lines for an hypothetical, unrealizable scenario, where the state B molecules are stable, i.e., $\gls{prSwitchSpon} = 0$. The results from Monte Carlo simulation are depicted by markers.}\label{fig:BER_over_velocity}
\end{figure}

In \Figure{fig:BER_over_velocity}, the \ac{BER} is shown as a function of flow velocity $v$ for two different \ac{EX} irradiation power densities and different spontaneous switching rate constants, which result in different values for $\gls{prSwitchSpon}$. Here, $\gls{prAEX} = 0.5$, $\gls{TTX} =  5 \times 10^{-5} \,\si{\second}$, $\gls{PowerRXin} = 2 \times 10^{12}\, \si{\watt \per\m \squared}$, $\gls{PowerTX} = 1 \times 10^{9}\, \si{\watt \per\m \squared}$, $\gls{zbTX} = 5 \times 10^{-3} \,\si{\meter}$, and $\gls{zbRX} = 0.405 \,\si{\meter}$ are used to satisfy all assumptions detailed in \Appendix{sections_assumptions}\footnote{We note that for very low flow velocity values, e.g., $v \leq 1\, \times 10^{-3}\,\si{\meter \per \second}$ (lower than in \Figure{fig:BER_over_velocity}), \Assumption{assump:spontaneous_switch} looses its validity.}.

First, we consider the case with spontaneous switching typical for \ac{GFPD} according to \Equation{eq:spontaneous_switching} (solid lines). We observe that the \ac{BER} has a unique minimum at a particular flow velocity, which we denote by $v_{\mathrm{opt}}$. Increasing and decreasing $v$ compared to $v_{\mathrm{opt}}$ increases the \ac{BER}. The reason is that a decrease in the flow velocity increases the time a molecule spends in the channel between \ac{EX} and \ac{RX}, which leads to an increase of \gls{prSwitchSpon} and a degradation of information transmission. In contrast, for high velocities, the conversion of state A molecules to state B molecules at \ac{EX} is insufficient to guarantee conversion of all molecules, as the irradiation duration $\gls{TEX}$ of a molecule inside the \ac{EX} is short. We also observe from \Figure{fig:BER_over_velocity} that an increase in $\gls{PowerEX}$ decreases the \ac{BER} if $v$ is large. In addition, $v_{\mathrm{opt}}$ is larger for higher $\gls{PowerEX}$.

Next, we consider $\gls{prSwitchSpon} = 0$.
In particular, $\gls{prSwitchSpon} = 0$ is an unrealizable special case for the proposed \ac{GFPD} as the half-time is not an adjustable property of the signaling molecules, cf. \Section{spSwitching}. However,
we consider this hypothetical case to gain insight into this specific property of media modulation based on \ac{GFPD}.

Compared to $\gls{prSwitchSpon} > 0$, for $\gls{prSwitchSpon} = 0$, the unique \ac{BER} minimum occurs at a smaller \ac{BER} and smaller velocity, which we denote as $v_{\mathrm{opt},0}$, i.e., $v_{\mathrm{opt},0} < v_{\mathrm{opt}}$. Decreasing $v$ from $v_{\mathrm{opt},0}$ increases the propagation duration from \ac{TX} to \ac{RX} and therefore increases the spatial spread of the modulated signaling molecules due to diffusion. Thus, $\mathbb{E}\{N_{\gls{PhotonOutRX}}\}$ for $s=1$ decreases, while it remains unchanged for $s=0$, which increases the \ac{BER}. For velocities larger than $v_{\mathrm{opt},0}$ the insufficiency at the \ac{EX} determines the performance and leads to a larger \ac{BER}. From the \ac{BER} difference between $\gls{prSwitchSpon} > 0$ and $\gls{prSwitchSpon} = 0$ in \Figure{fig:BER_over_velocity} we conclude that for small velocities $\gls{prSwitchSpon} = 0$ is beneficial. These insights motivate the design of new photochromic molecules, whose states are more stable, i.e., which have a smaller spontaneous switching rate constant.

\scaleSection
\section{Conclusion}
\label{sec:conclusion}
\scaleSectionBelow
In this paper, we introduced a new form of media modulation for \ac{MC}, which provides a promising perspective for designing non-invasive MC systems. Media modulation does not require a \ac{TX} that stores signaling molecules and controls their release. In particular, in media modulation based \ac{MC} the \ac{TX} only alters the state of the signaling molecules already present in the channel. We investigated the properties of media modulation for the group of photochromic molecules, whose states can be controlled by external light stimuli. This control mechanism allows for writing, reading, and erasing of information embedded in the molecule state. Furthermore, we studied the usage of these molecules for information transmission in a \ac{3D} duct system with one \ac{EX}, one \ac{TX}, and one \ac{RX}. We developed a statistical model for the received signal taking into account eight consecutive and interdependent random processes. Based on the proposed statistical model, we derived analytical expressions for the optimal threshold value of a threshold based detector and the \ac{BER}. Our numerical \ac{BER} results prove that even in the absence of a molecule emitting \ac{TX} reliable information transmission is feasible. Furthermore, our results reveal that (i) increasing the \ac{EX}, \ac{TX}, and \ac{RX} power densities leads to a \ac{BER} reduction, (ii) media modulation is negatively affected by \ac{TX} noise and spontaneous switching in the channel, and (iii) there exists an optimal flow velocity due to the tradeoff between reliable switching at the \ac{EX} for low velocity values and low spontaneous switching as well as reduced molecule spread in the channel for high velocity values.

In this paper, we focused on media modulation based on the photochromic molecule \ac{GFPD}. A detailed investigation of signaling molecules which can be switched to more than two states and therefore allow for higher order modulation is an interesting topic for future work. Furthermore, it would be interesting to analyze the applicability of other biological processes, such as phosphorylation, for synthetic media modulation based \ac{MC}. Moreover, the investigation of potential applications of media modulation in industrial pipelines, bio-medical environments, bio-chemical reactors, and microfluidic testbeds seems promising. Also, while a general fair comparison between media modulation and conventional modulation schemes does not seem possible because of the different transmitter and system models, performance and complexity comparisons for concrete application scenarios are important. In addition, while we assumed uniform flow for simplicity of presentation, the extension of the proposed designs and models to laminar, turbulent, and time-varying flows is of high theoretical and practical interest.

\scaleSection
\appendix
\section{}
\scaleSubsection
\subsection{Collection of Simplifying Assumptions}\label{sections_assumptions}
\scaleSubsectionBelow
Here, we summarize the assumption made in \Sections{photoChemic}{sec:math}.
\begin{assump}
\textit{All signaling molecules are exposed to the irradiation inside \ac{EX} for $\gls{TEX} = \frac{\gls{lengthEX}}{v}$ seconds.}
Given the assumed uniform flow, this assumption holds if the molecule movement due to diffusion within \ac{EX} is negligible, i.e., for large P\'{e}clet numbers \cite[Eq. (20)]{Jamali2019ChannelMF} with
$
  \mathrm{Pe} = \frac{\gls{lengthEX} \,v}{D} > \chi \gg 1.
$
Here, for convenience, $\chi$ denotes the threshold that we considered to guarantee $\gg 1$ in the evaluation part in \Section{sec:evaluation}.
\label{assumptionEX}
\end{assump}

\begin{assump}
  \textit{During one irradiation of length $\gls{TTX}$ molecules do not move into or out of $\gls{VolumeTX}$, i.e., they are static, and are equally effected by the irradiation.}
The assumption is valid if the random displacement of a molecule within the irradiation duration $\gls{TTX}$ is much shorter than the \ac{TX} length $\gls{lengthTX}$, i.e.,
$
  \frac{\gls{lengthTX}}{\underbrace{\sqrt{2 D \, \gls{TTX}}}_{\mathrm{Diffusion}} + \underbrace{v \, \gls{TTX}}_{\mathrm{Flow}}} > \chi \gg 1  \;,
$
where the propagation distance due to diffusion is provided by its standard deviation $\sqrt{2 D \gls{TTX}}$.
\label{cond_static_TX}
\end{assump}

\begin{assump}
  \textit{Spontaneous switching mainly happens in the propagation channel from \ac{TX} to \ac{RX}.}
  This assumption holds true, if
$
  \frac{T_{1/2}}{T^{\scriptscriptstyle{\mathrm{Ch}}}(\gls{zaTX}, \gls{zbTX})}  = \frac{T_{1/2}}{\frac{ \gls{lengthTX}}{v}} > \chi \gg 1 \quad \land \quad \frac{T_{1/2}}{T^{\scriptscriptstyle{\mathrm{Ch}}}(\gls{zaEX}, \gls{zbEX})}  = \frac{T_{1/2}}{\frac{ \gls{lengthEX}}{v}} > \chi \gg 1\;,
$
as in this case $p_{\mathrm{s,SP}}(\gls{zaTX}, \gls{zbTX}) \approx 0 $ and $p_{\mathrm{s,SP}}(\gls{zaEX}, \gls{zbEX}) \approx 0 $ follows.
\label{assump:spontaneous_switch}
\end{assump}

\begin{assump}
  \textit{The molecules are static during the fluorescence process of duration \gls{TRX}, which starts at $t_\mathrm{s}$.}
  This assumption is valid, if the duration \gls{TRX} to read out the transmitted information at the \ac{RX} is short, i.e.,
$
    \frac{\gls{lengthRX}}{\underbrace{\sqrt{2 D \, \gls{TRX}}}_{\mathrm{Diffusion}} + \underbrace{v \, \gls{TRX}}_{\mathrm{Flow}}} > \chi \gg 1  \;,
$ cf. condition for \Assumption{cond_static_TX}.
  \label{RX_static}
\end{assump}

\begin{assump}
  \textit{$t \in [0, 0 + \gls{TTX}]$ is approximated by $t=0$.}
  This approximation is valid for small $\gls{TTX}$, i.e., when no molecules enter or leave the \ac{TX} volume $\gls{VolumeTX}$ during the modulation time length.
  \label{assump_modulation_time}
\end{assump}
\scaleSubsection
\subsection{Number of Switched Molecules}\label{section:appendix:switching}
\scaleSubsectionBelow
Eq. \Equation{eq:beer_concentration} can be solved for $C_{X}(t)$ as follows:
\scaleAlign
\begin{align}
   \Equation{eq:beer_concentration} \overset{(a)}{\Leftrightarrow} \, &  \frac{P_{\mathrm{in}}^{m}}{  E_{\mathrm{in}}^{m} H  N_{\mathrm{Av}}} \,\varphi_{X} \, \mathrm{d}t = - \frac{1}{1-\exp\big(-a C_{X}(t)\big)}\mathrm{d}C_{X}(t) \nonumber \\
  \overset{(b)}{\Leftrightarrow} \, &  \frac{P_{\mathrm{in}}^{m}}{  E_{\mathrm{in}}^{m} H  N_{\mathrm{Av}}} \,\varphi_{X} \, t + \mathrm{c} = - \frac{\log\big(1- \exp( a C_{X}(t))\big)}{a}\nonumber \\
  \overset{(c)}{\Leftrightarrow} \, & \exp\left(-a \,  \frac{P_{\mathrm{in}}^{m}}{  E_{\mathrm{in}}^{m} H  N_{\mathrm{Av}}} \,\varphi_{X} \, t\right) \big(1-\exp(a C_{{X}, 0})\big) \nonumber \\
  & \qquad = 1- \exp( a C_{X}(t)) \;,
\end{align}
where we apply in $(a)$ the substitution $a \mkern-2mu= \mkern-2mu\log(10) H \epsilon^{m}$ and multiplication with $\mathrm{d}t$ and in $(b)$ the integration \ac{w.r.t.} $t$ and $C_{X}(t)$. Subsequently, we exploit in $(c)$ $C_{{X}}(t=0) \mkern-2mu=\mkern-2mu C_{{X}, 0}$, which yields $\mathrm{c} \mkern-2mu=\mkern-2mu - \frac{\log\big(1- \exp( a C_{{X}, 0})\big)}{a}$. Then, we apply the exponential function to both sides of the equation. Finally, after further rearranging, we apply the \mbox{natural logarithm function to both sides to obtain \Equation{diff_solution}.}
\scaleSubsection
\subsection{Derivation of Minimum Illumination Time at RX to Reach Equilibrium}\label{Equilibrium_state_time}
\scaleSubsectionBelow
As \Equation{excited_equilibrium_number} denotes the steady-state concentration and does not contain the transient part, we derive the minimum illumination duration $T_{\mathrm{eq,min}}$ necessary to reach the steady state, i.e., the time required for \Equation{excited_equilibrium_number} to be valid. To improve readability, we change variables as $t' = t-t_\mathrm{s}$. As the solution to the non-linear differential equation defined by \Equations{fluorescence}{beer_lambert_rule_fluorescence} is not straightforward, we approximate the exponential function in \Equation{beer_lambert_rule_fluorescence} by the first two terms of its Taylor series, i.e., $\exp(x) \approx 1+x$, which leads to
\scaleAlign
\begin{align}
  \qquad &\frac{\mathrm{d}C_{\mathrm{A}^{\mathrm{G}}}^{\scriptscriptstyle{\mathrm{RX}}}(t')}{\mathrm{d}t'} = (k_{\mathrm{r}} + k_{\mathrm{nr}}) C_{\mathrm{A}^*}^{\scriptscriptstyle{\mathrm{RX}}} - \frac{\gls{PowerRXin} \log(10) \gls{epsilonRX}}{  \gls{EnergyPhotonRX}  N_{\mathrm{Av}}} C_{\mathrm{A}^{\mathrm{G}}}^{\scriptscriptstyle{\mathrm{RX}}}(t') \nonumber \\
  \overset{(a)}{\Rightarrow}\, & \frac{\mathrm{d}C_{\mathrm{A}^{\mathrm{G}}}^{\scriptscriptstyle{\mathrm{RX}}}(t')}{\mathrm{d}t'} + z_0 C_{\mathrm{A}^{\mathrm{G}}}^{\scriptscriptstyle{\mathrm{RX}}}(t') = C_{\mathrm{A}}^{\scriptscriptstyle{\mathrm{RX}}} z_1 \nonumber \\
  \overset{(b)}{\Rightarrow}\, & C_{h,\mathrm{A}^{\mathrm{G}}}^{\scriptscriptstyle{\mathrm{RX}}}(t') = c_0 \exp(-z_0 t')\nonumber \\
  \overset{(c)}{\Rightarrow}\, & \frac{\mathrm{d}c(t')}{\mathrm{d}t'} = C_{\mathrm{A}}^{\scriptscriptstyle{\mathrm{RX}}} z_1 \exp(z_0 t') \nonumber \\
  \overset{(d)}{\Rightarrow}\, & C_{\mathrm{A}^{\mathrm{G}}}^{\scriptscriptstyle{\mathrm{RX}}}(t')  = \exp(-z_0 t') c_1 + \frac{C_{\mathrm{A}}^{\scriptscriptstyle{\mathrm{RX}}} z_1}{z_0}\nonumber \\
  \overset{(e)}{\Rightarrow}\, & C_{\mathrm{A}^{\mathrm{G}}}^{\scriptscriptstyle{\mathrm{RX}}}(t') = C_{\mathrm{A}}^{\scriptscriptstyle{\mathrm{RX}}} \left[ \exp(-z_0 t') - \exp(-z_0 t') \frac{z_1}{z_0} + \frac{z_1}{z_0} \right] ,
  \label{linear_DGL_fluorescence}
\end{align}
for $ t'\geq 0 $, where we substitute in $(a)$ $z_0 = k_{\mathrm{r}} + k_{\mathrm{nr}} + \frac{\gls{PowerRXin} \log(10) \gls{epsilonRX}}{  \gls{EnergyPhotonRX}  N_{\mathrm{Av}}}$, $z_1= k_{\mathrm{r}} + k_{\mathrm{nr}}$, and $C_{\mathrm{A}^*}^{\scriptscriptstyle{\mathrm{RX}}}(t') = C_{\mathrm{A}}^{\scriptscriptstyle{\mathrm{RX}}} - C_{\mathrm{A}^{\mathrm{G}}}^{\scriptscriptstyle{\mathrm{RX}}}(t')$, provide in $(b)$ the homogeneous solution with $c_0 \in \mathbb{R}$, and utilize in $(c)$ the variation of constants, which yields the general solution in $(d)$ with $c_1 \in \mathbb{R}$. Finally, we exploit in $(e)$ that all state A molecules are in substate $\mathrm{A}^{\mathrm{G}}$ at the beginning of the fluorescence process, i.e., $C_{\mathrm{A}^{\mathrm{G}}}^{\scriptscriptstyle{\mathrm{RX}}}(t'=0) = C_{\mathrm{A}}^{\scriptscriptstyle{\mathrm{RX}}}$.
We assume the equilibrium state to be reached, if
$
  \Big|\frac{\mathrm{d}C_{\mathrm{A}^{\mathrm{G}}}^{\scriptscriptstyle{\mathrm{RX}}}(t')}{\mathrm{d}t'}\Big|_{t'= T_{\mathrm{eq,min}}} < \frac{\epsilon_{\mathrm{eq}}}{N_{\mathrm{Av}}\gls{VolumeSizeRX}} \;,
$
with precision error $\epsilon_{\mathrm{eq}}$, which yields
\scaleAlign
\begin{align}
  T_{\mathrm{eq,min}} \approx  - \ln\left(\frac{\epsilon_{\mathrm{eq}}}{C_{\mathrm{A}}^{\scriptscriptstyle{\mathrm{RX}}}N_{\mathrm{Av}}\gls{VolumeSizeRX}(z_0-z_1)}\right) \frac{1}{z_0} \;.
  \label{eq:Teq_min}
\end{align}
\scaleSubsection
\subsection{Derivation of the Fluorescence Intensity}\label{section:appendix:fluorescence}
\scaleSubsectionBelow
We insert \Equation{beer_lambert_rule_fluorescence} in \Equation{fluorescence} and substitute \mbox{$x_0\mkern-2mu = \mkern-2mu\log(10) \gls{epsilonRX} \, \mkern-3mu H $, which in conjunction with \Equation{equilibrium_fluorescence} yields}
\scaleAlign
\begin{align}
    &(k_{\mathrm{r}} + k_{\mathrm{nr}}) C_{\mathrm{A}^*,\,\mathrm{eq}}^{\scriptscriptstyle{\mathrm{RX}}} -  \frac{\gls{PowerRXin}}{  \gls{EnergyPhotonRX} H  N_{\mathrm{Av}}} \mkern-3mu\bigg(\mkern-3mu1-\exp\mkern-3mu\big(\mkern-3mu-x_0 C_{\mathrm{A}^{\mathrm{G}},\,\mathrm{eq}}^{\scriptscriptstyle{\mathrm{RX}}}\big)\mkern-5mu\bigg)\mkern-3mu = 0 \nonumber \\
    &\overset{(a)}{\Rightarrow} \, x_1 C_{\mathrm{A}^*,\,\mathrm{eq}}^{\scriptscriptstyle{\mathrm{RX}}} = 1 - x_2 \exp\big(x_0 C_{\mathrm{A}^*,\,\mathrm{eq}}^{\scriptscriptstyle{\mathrm{RX}}}\big) \nonumber \\
    &\overset{(b)}{\Rightarrow} \, x_0 C_{\mathrm{A}^*,\,\mathrm{eq}}^{\scriptscriptstyle{\mathrm{RX}}} \exp\left(\frac{x_0}{x_1}-x_0 C_{\mathrm{A}^*,\,\mathrm{eq}}^{\scriptscriptstyle{\mathrm{RX}}}\right) \nonumber \\
    & \qquad = \frac{x_0}{x_1} \exp\left(\frac{x_0}{x_1}-x_0 C_{\mathrm{A}^*,\,\mathrm{eq}}^{\scriptscriptstyle{\mathrm{RX}}}\right) - \frac{x_0 x_2}{x_1} \exp\left(\frac{x_0}{x_1}\right) \nonumber \\
    &\overset{(c)}{\Rightarrow} \,  \frac{x_0 x_2}{x_1} \exp\left(\frac{x_0}{x_1}\right) \nonumber \\
    &\qquad = \left(\frac{x_0}{x_1}- x_0 C_{\mathrm{A}^*,\,\mathrm{eq}}^{\scriptscriptstyle{\mathrm{RX}}}\right) \exp\left(\frac{x_0}{x_1}-x_0 C_{\mathrm{A}^*,\,\mathrm{eq}}^{\scriptscriptstyle{\mathrm{RX}}}\right) \nonumber \\
    &\overset{(d)}{\Rightarrow} \,  \frac{x_0}{x_1}- x_0 C_{\mathrm{A}^*,\,\mathrm{eq}}^{\scriptscriptstyle{\mathrm{RX}}} = W\left(\frac{x_0 x_2}{x_1} \exp\left(\frac{x_0}{x_1}\right)\right) \;,
\end{align}
where we substituted in $(a)$ $x_1 = \frac{ \gls{EnergyPhotonRX} H N_{\mathrm{Av}} (k_{\mathrm{r}} + k_{\mathrm{nr}})}{\gls{PowerRXin}}$ and $x_2 = \exp(-x_0 C_{\mathrm{A}}^{\scriptscriptstyle{\mathrm{RX}}})$, and used the classification of signaling molecules into substates \Equation{N_A_subversions} by $C_{\mathrm{A}^{\mathrm{G}},\,\mathrm{eq}}^{\scriptscriptstyle{\mathrm{RX}}} =  C_{\mathrm{A}}^{\scriptscriptstyle{\mathrm{RX}}}- C_{\mathrm{A}^*,\,\mathrm{eq}}^{\scriptscriptstyle{\mathrm{RX}}}$. Subsequently, we apply in $(b)$ an expansion by factor $\frac{x_0}{x_1} \exp\left(\frac{x_0}{x_1}\right) \exp\left(-x_0 C_{\mathrm{A}^*,\,\mathrm{eq}}^{\scriptscriptstyle{\mathrm{RX}}}\right)$. Furthermore, after factoring in $(c)$, we substitute in $(d)$ the Lambert W function $W(\cdot)$ defined as $w \exp(w) = z \Rightarrow W(z) = w$. Finally, further rearranging yields \Equation{excited_equilibrium_number}.

\scaleSubsection
\subsection{Derivation of Impulse Responses}\label{section:appendix:IR}
\scaleSubsectionBelow
Because of the assumed uniform flow and the transparent, counting \ac{RX}, which covers the entire cross-section of the pipe, the received signaling molecule concentration is independent of the $xy$ dimension. Thus, the considered \ac{3D} environment can be equivalently modeled by an unbounded \ac{1D} environment. The \ac{1D} environment is unbounded as the pipe has infinite axial extent. The expected spatio-temporal molecule concentration $P$ in the equivalent \ac{1D} model is characterized by the advection-diffusion equation given as $\partial_{t} P = D \partial_z^2 P - v \partial_z P$. The solution to this equation is given by $P(t, z^{\scriptscriptstyle{\mathrm{RX}}}, z^{\scriptscriptstyle{\mathrm{TX}}}) = \frac{1}{\sqrt{4 \pi D \, t}} \exp\left( - \frac{(z^{\scriptscriptstyle{\mathrm{RX}}}-z^{\scriptscriptstyle{\mathrm{TX}}} - v t)^2}{4 D \, t}\right)$, i.e., $P(t, z^{\scriptscriptstyle{\mathrm{RX}}}, z^{\scriptscriptstyle{\mathrm{TX}}})$ denotes the probability that one molecule released at $z^{\scriptscriptstyle{\mathrm{TX}}}$ at time $t=0$ is observed at $z^{\scriptscriptstyle{\mathrm{RX}}}$ at time $t > 0$.

The probability of a signaling molecule to be within the \ac{RX} is obtained by expecting over the release position $z^{\scriptscriptstyle{\mathrm{TX}}}$ and integration over the axial extent of the \ac{RX} as follows
\scaleAlign
  \begin{align}
  h(t) &= \underset{z^{\scriptscriptstyle{\mathrm{TX}}}}{\mathbb{E}} \left\{ \int \limits_{\gls{zaRX}}^{\gls{zbRX}} P(t, z^{\scriptscriptstyle{\mathrm{RX}}}, z^{\scriptscriptstyle{\mathrm{TX}}}) \dzr \right\} \nonumber \\
  &=\int \limits_{\gls{zaTX}}^{\gls{zbTX}} \int \limits_{\gls{zaRX}}^{\gls{zbRX}} P(t, z^{\scriptscriptstyle{\mathrm{RX}}}, z^{\scriptscriptstyle{\mathrm{TX}}}) \textnormal{f}_{z^{\scriptscriptstyle{\mathrm{TX}}}}\left(z^{\scriptscriptstyle{\mathrm{TX}}}\right) \dzr \dzm \nonumber \\
  &\overset{{(a)}}{=} \frac{1}{2 \,\gls{lengthTX}}\int \limits_{\gls{zaTX}}^{\gls{zbTX}} \erf\left( \frac{\gls{zbRX} - z^{\scriptscriptstyle{\mathrm{TX}}} - v t}{\sqrt{4 D \, t}}\right) \nonumber \\
  &\qquad - \erf\left( \frac{\gls{zaRX} - z^{\scriptscriptstyle{\mathrm{TX}}} - v t}{\sqrt{4 D \, t}}\right) \dzm \,,
\end{align}
  where $\mathbb{E}\{\cdot \}$ denotes the expectation operator, and we exploit in $(a)$ the uniform distribution of $z^{\scriptscriptstyle{\mathrm{TX}}}$, i.e., $\textnormal{f}_{z^{\scriptscriptstyle{\mathrm{TX}}}}\left(z^{\scriptscriptstyle{\mathrm{TX}}}\right) = \frac{1}{\gls{lengthTX}} $, $\gls{zaTX} \leq z^{\scriptscriptstyle{\mathrm{TX}}} \leq \gls{zbTX}$. Finally, applying the integral $\int \erf(x) \dx = x \erf(x) + \frac{1}{\sqrt{\pi}} \exp\left( -x^2\right)$, we obtain \Equation{cir_analytic}. This concludes the proof.

\scaleSubsection
\subsection{Distribution of a Doubly Stochastic Binomial Process}\label{section:appendix:conditionalProbabilities}
\scaleSubsectionBelow
\begin{theo}
  If $F$ and $G$ given $F$ are Binomial distributed random variables, then $G$ again follows a Binomial distribution, i.e.,
  \scaleAlign
  \begin{align}
    G &\sim \binomial{n, p \, q} \;, \nonumber \\
     &\quad \mathrm{if} \quad G \given F \sim \binomial{F, q} \wedge F \sim \binomial{n, p} \;,
    \label{eq:condBinom}
  \end{align}
  where $\binomial{N, p}$ denotes a binomial distribution with parameters $N$ and $p$. Here, $N$ and $p$ denote the number of trials and the success probability, respectively.
\end{theo}
\begin{IEEEproof}
  As $F \sim \binomial{n, p}$ and $G \given F \sim \binomial{F, q}$, the total probability of $G$ is obtained as follows
  \scaleAlign
  \begin{align}
    &\prob{G = m} \nonumber \\
    &= \sum \limits_{k=m}^{n} \prob{G = m \given F = k} \prob{F = k} \nonumber \\
     &= \sum \limits_{k=m}^{n} \binom{k}{m} q^{m} (1-q)^{k-m} \binom{n}{k} p^{k} (1-p)^{n-k} \nonumber \\
     &\overset{(a)}{=} \sum \limits_{k=m}^{n} \binom{n}{m} \binom{n-m}{k-m} q^{m} (1-q)^{k-m}  p^{k} (1-p)^{n-k} \nonumber \\
     &\overset{(b)}{=} \binom{n}{m} (q p)^{m} \sum \limits_{i=0}^{n-m} \binom{n-m}{i} (p-q p)^i (1-p)^{N-i-m}  \nonumber \\
     &\overset{(c)}{=} \binom{n}{m} (q p)^{m} (1-q p)^{n-m}  \;,
  \end{align}
  where in $(a)$ we exploit $\binom{k}{m}  \binom{n}{k} \mkern-2mu=\mkern-2mu \binom{n}{m} \binom{n-m}{k-m}$, in $(b)$ we use $p^k \mkern-2mu=\mkern-2mu p^{k-m} p^{m}$ and substitute $i \mkern-2mu=\mkern-2mu k-m$, and in $(c)$ we use the binomial \mbox{identity $(a+b)^c \mkern-2mu=\mkern-2mu \sum \limits_{d=0}^{c} \binom{c}{d} a^{c-d} b^{d}$. This completes the proof.}
\end{IEEEproof}
\bibliographystyle{IEEEtran}
\bibliography{literature}
\end{document}